\documentclass[12pt,notoc]{JHEP3}

\usepackage{amsmath,amssymb,euscript,array,cite,mathrsfs,amsfonts}
\usepackage{epsfig}

\newcommand{\startappendix}{
\setcounter{section}{0}
\renewcommand{\thesection}{\Alph{section}}}
\newcommand{\Appendix}[1]{
\refstepcounter{section}
\begin{flushleft}
{\large\bf Appendix \thesection: #1}
\end{flushleft}}

\newcommand{\field}[1]{\mathbb{#1}} 
\newcommand*{\Dsl}[0]{{\rlap{\kern2.25pt /}{D}}}
\newcommand*{\Asl}[0]{{\rlap{\kern2.25pt /}{A}}}
\newcommand*{\dsl}[0]{{\rlap{\kern0.5pt /}{\partial}}}
\newcommand*{\xisl}[0]{{\rlap{\kern0.5pt /}{\xi}}}
\newcommand*{\asl}[0]{{\rlap{\kern0.5pt /}{a}}}
\newcommand*{\bsl}[0]{{\rlap{\kern0.5pt /}{b}}}

\newcommand*{\Tr}[0]{{\rm Tr}}

\def\Dslash{\,\,{\raise.15ex\hbox{/}\mkern-12mu D}}

\newcommand{\EQ}[1]{\begin{equation} #1 \end{equation}}
\newcommand{\SP}[1]{\begin{equation}\begin{split} #1
\end{split}\end{equation}}
\def\BZ{{\mathbb Z}}
\def\Z{{\mathbb Z}}
\def\R{{\mathbb R}}

\title{Finite Volume Phases of Large-${\boldsymbol N}$ Gauge Theories 
with Massive Adjoint Fermions}
\author{Timothy J. Hollowood and Joyce C. Myers\\ Swansea University,
  Physics Department, Swansea SA2 8PP, UK\\ E-mail:
  \email{t.hollowood@swansea.ac.uk, j.c.myers@swan.ac.uk}}

\abstract{The phase structure of QCD-like gauge theories with fermions
  in various representations is an interesting but generally
  analytically intractable problem. One
  way to ensure weak coupling is to define the theory in
  a small finite volume, in this case $S^3\times S^1$. Genuine phase
  transitions can then occur in the large $N$ theory. Here, we use this
  technique to investigate $SU(N)$ gauge theory with a number $N_f $ of
  massive adjoint-valued Majorana fermions having non-thermal boundary
  conditions around $S^1$. For $N_f =1$ we find a line of
  transitions that separate the weak-coupling analogues of the
  confined and de-confined phases for which the density of
  eigenvalues of the Wilson line 
transform from the uniform distribution to a
  gapped distribution. However, the situation for $N_f >1$
  is much richer and a series of weak-coupling analogues of
  partially-confined phases appear which leave unbroken a $\Z_p$
  subgroup of the centre symmetry. In these $\Z_p$ phases the
  eigenvalue density has $p$ gaps and they are separated from the
  confining phase and from
  one-another by first order phase transitions.  
We show that for small enough $mR$ 
(the mass of the fermions times the radius of the $S^3$) only the
  confined phase exists.
The large $N$ phase diagram is consistent with the finite $N$ result and with other
  approaches based on $\R^3\times S^1$ calculations 
and lattice simulations.
} 


\preprint{}

\begin{document}


\section{Introduction}

The study of gauge theories in finite volume, and specifically
$S^3 \times S^1$, has been an interesting and fruitful one. A weak coupling
regime is ensured by taking the size of the compact space to be small
compared with the strong coupling scale, ${\rm min}[R_{S^1}, R_{S^3}] \ll \Lambda_{QCD}^{-1}$.
The theory is then non-trivial even at the one-loop level
because the projection onto gauge invariant states introduces
effective interactions between the gluons 
\cite{Aharony:2003sx,Aharony:2005bq,Hallin:1998km,Sundborg:1999ue}. The large $N$
limit is taken in order to ensure that a thermodynamic limit exists
and genuine phase transitions occur. One
motivation is to study thermal properties of gauge theories, in which
case the $S^1$ is interpreted as the ``thermal circle'' and fermions
have anti-periodic boundary conditions. A
one-loop calculation is then sufficient to
uncover the weak-coupling manifestation of the
confinement/de-confinement transition in that on one side of the
transition the expectation value of the Polyakov loop vanishes---the
confining phase---while
on the other side---the de-confined phase---it gains a VEV. The
transition is Hagedorn-like in that the density of states grows
exponentially in the low temperature phase. In order
to ascertain the order of the transition 
higher loop effects are crucial \cite{Aharony:2005bq}. In pure gauge
theory it is known to be a first order transition that occurs at a
lower temperature than the Hagedorn transition in the non-interacting theory. 
The same kind of transition occurs
when adjoint matter fields are added, and in particular for the ${\cal N}=4$
gauge theory. One of the deep insights to emerge from the AdS/CFT
correspondence is that this transition can also be seen in the strong
coupling gravity
dual as a Hawking-Page transition from thermal AdS space to an AdS
black hole \cite{Witten:1998zw}.
In the present paper, since we are interested in theories
with periodic boundary conditions for fermions around the $S^1$ the
phase transitions are quantum rather than thermal and the connection
with the AdS/CFT correspondence is not so obvious even though one
could imagine obtaining our theories from an ${\cal N}=4$ theory with SUSY
breaking mass deformations.

The phase diagram of $SU(N)$ gauge theories in finite volume can be
studied in other ways. At strong coupling lattice simulations are
the dominant technique for obtaining the phase diagram. 
The phase diagrams using the strong and weak coupling  techniques have
so far not been easy to compare, since the relationship between
lattice bare parameters and continuum renormalized parameters is not
clear. However, in some cases a qualitative comparison is
possible. In this paper we explore this possibility in adjoint QCD, that is,
$SU(N)$ gauge theory with fermions in the adjoint
representation. The phase diagram of this theory is quite rich when
considering $N_f>1$ Majorana flavours with fermions of finite mass to which periodic boundary conditions have been applied in the temporal direction
\cite{Myers:2007vc,Wozar:2008nv,Myers:2009df,Bedaque:2009md,Cossu:2009sq,Bringoltz:2009mi,Bringoltz:2009kb}.
The intuition behind this is that the gauge field and fermion terms in the effective potential have opposite signs and compete
to dominate the Polykov loop, with the fermions having a disordering
effect. For a sufficient number of light fermions, the disordering
effect dominantes and a confining phase results with vanishing Polakov
loop, however as the
masses are increased the disordering effect becomes weaker and a phase
transition can occur where the centre symmetry is (partially) broken.

The issues that we investigate in this paper 
are relevant to some active areas of
research. The application of periodic boundary conditions to adjoint
representation fermions causes the confined phase to be accessible at
weak coupling \cite{Kovtun:2007py,Myers:2007vc,Myers:2009df} \footnote{It should be noted that even though the confined phase is perturbatively accessible in adjoint QCD when ${\rm min}[R_{S^1}, R_{S^3}] \ll \Lambda^{-1}$, it was found in \cite{Unsal:2008ch} that semiclassical analyses on ${\field R}^3 \times S^1$, specifically including the contribution of magnetically charged objects in the confined phase, are only valid when $R_{S^1} N \Lambda \ll 1$. For certain observables which exhibit volume independence in the confined phase this should also be true on $S^3 \times S^1$. We thank Mithat Unsal for pointing out this important result.}. This is
interesting for several reasons. For one, it is possible that observables may not differ significantly in the pure Yang-Mills theory confined phase and the
perturbative confined phase of adjoint QCD. The latter feature might
na\"\i vely be inferred from the results of lattice calculations of the
string tensions in the confining phase of the pure Yang-Mills theory
which show little temperature dependence. However, it is also known
that the string tensions have a significant temperature dependence
above $T_c$ in the de-confined phase \cite{Boyd:1996bx}. This suggests
an important question: Is the temperature dependence a result of the
change in confinement scale when moving above $T_c$ such that it would
also occur in a high temperature confined phase, or is it something
intrinsic to the de-confined phase? In the case where the high
temperature confined phase is induced by periodic boundary conditions
the question turns into one of dependence of observables on the length
$L$ of $S^1$ in the weak-coupling confined phase. In this case
the question of temperature dependence becomes one of volume dependence. 

The idea of volume dependence---or rather independence---is particularly
interesting in the context of large $N$ confining theories.
In 1982
Eguchi and Kawai proposed volume independence in large $N$ Yang-Mills
theory in \cite{Eguchi:1982nm} where they employed large $N$
factorization to show that pure Yang-Mills theory formulated on a
lattice at some arbitrary volume, can be mapped onto the theory
formulated on a single site. Around that time it was also shown that
volume independence can only hold if certain symmetries are not
broken, in particular it can only hold in the confining phase of large
$N$ gauge theories \cite{Yaffe:1981vf,Bhanot:1982sh,Kazakov:1982gh,Okawa:1982ic}. Several ideas were proposed to maintain the ${\field Z}_N^d$ symmetry which is required on a space with $d$ independently compactified dimensions (i.e. where certain dimensions ${\field R}$ are compactified to $S^1$). Two of the most promising proposals were the quenched \cite{Bhanot:1982sh}, and twisted \cite{GonzalezArroyo:1982ub} Eguchi-Kawai models. In the quenched EK model the eigenvalues of the Polyakov loop quenched, that is their eigenvalues are set by hand, such that ${\field Z}_N^d$ symmetry required is explicit. However, results from lattice simulations of $SU(N)$ gauge theory in \cite{Bringoltz:2008av} show that large $N$ reduction using the quenched EK model breaks down. In the twisted EK model the boundary conditions are twisted by multiplying each plaquette in the lattice action by an element of the center such that the action becomes invariant under the required ${\field Z}_N^d$ symmetry. The twisted EK model was shown to break down in \cite{Teper:2006sp,Azeyanagi:2007su}. More
recently, the idea of large $N$ volume independence was picked up
again and generalized in \cite{Kovtun:2007py} where the authors
proposed that since QCD(Adj) has a confined phase which is
perturbatively accessible, a generalized Eguchi-Kawai large $N$
reduction could relate the weak-coupling, small volume confined phase,
to the strong coupling, large volume confined phase. Since this
proposal there have been several tests. In \cite{Bedaque:2009md} the authors performed a weak-coupling
calculation of the effective potential for a three-dimensional adjoint
matter theory which is related by generalized orbifold projection to
QCD(Adj) in four dimensions with one dimension compact. In
\cite{Bringoltz:2009mi} it was shown that this calculation is
renormalization scheme dependent and that for Eguchi-Kawai reduction
to hold it requires double trace counter-terms with coefficients
defined so that the $\Z_N$ symmetry is preserved in the limit  of zero
adjoint fermion mass. In a lattice simulation \cite{Cossu:2009sq} for
$N = 3$ the authors calculated the phase diagram of QCD(Adj) on a
$16^3 \times L_c$ lattice and showed that there is a confined phase at
both strong and weak coupling but that they are separated by phase
transitions into other phases depending on the value of the adjoint
fermion mass. From their results the presence, or lack thereof, of
phase transitions in the chiral limit is unclear. But, their result
places boundaries on the validity of large $N$ reduction either way
since it can only hold when the theories to be mapped are in confining
phases. Most recently in \cite{Bringoltz:2009kb} the Eguchi-Kawai
reduced (single-site) model with dynamical adjoint fermions was
studied on the lattice and it was shown that the $\Z_N$ symmetry is
unbroken for light enough fermion mass. Some applications of Eguchi-Kawai reduction can be found in \cite{Hanada:2009kz,Ishiki:2009sg} where the authors use large $N$ reduction of SYM (${\cal N} = 1$ in \cite{Hanada:2009kz} and ${\cal N} = 4$ in \cite{Ishiki:2009sg}) on $S^3 \times {\field R}$ to reduce the theory to a single dimension for the purpose of studying supersymmetric matrix quantum mechanics.

Our calculations in this paper
place boundaries on validity of large-$N$ reduction at weak coupling and finite fermion mass by mapping the regions of $Z(N)$ symmetry breaking in the phase
diagram of large $N$ adjoint QCD. The phase diagram is calculated as a function of the length $L$ of $S^1$,
the radius $R$ of $S^3$, and the adjoint fermion mass $m$.  In
particular, for small $L/R$ the mass of the fermions, times the radius $R$, must be below a
critical value to keep the theory in a $Z(N)$ symmetric phase. It is clearly an important question to
understand the phases of QCD with adjoint fermions as a function of
the volume and mass and it is to this question that we now turn.

In Section 2 we compute the effective action for the theory on
$S^3\times S^1$ as a function of the Polyakov loop 
to one-loop order 
paying particular attention to the
inclusion of a mass for the fermions. Section 3 investigates the
phase diagram of the large $N$ theory as a function of the radii of
$S^3\times S^1$ and the mass of the fermions and for different numbers
of adjoint fermion flavours. Here we show the existence of a rich
phase structure for $N_f >1$.
In the final section, we consider the same theories with $N$ finite where the
phase transitions are no longer non-analytic but
allow comparisons with  
with earlier work mentioned above for the theory on $\R^3\times
S^1$ and lattice simulations.

\section{The Effective Action on $S^3\times S^1$}

In this section, we review the way that the effective action is
calculated on  $S^3\times S^1$ to the level of the one-loop
approximation. The only new ingredient over earlier work 
is the inclusion of a mass for
the fermions.
Our approach follows closely the philosophy set out in the 
beautiful paper \cite{Aharony:2003sx}, however, we shall use a 
more conventional form of gauge fixing, described in
\cite{Hollowood:2006cq,Hollowood:2006xb,Hollowood:2008gp}, 
that leads to the same result. 

We shall start with $SU(N)$ gauge theory with a number of Majorana
fermions $\psi_f$ transforming in representations $R_f$ 
of the gauge group.\footnote{In general in order to have a mass term
the representations $R_f$ must be real or include complex conjugate pairs.
In the case here we are considering the
adjoint representation which is real.} The action is
\EQ{
S=\frac1{g_{YM}^2}\int
d^4x\sqrt{g}\,\Big\{-\tfrac14\Tr\,F_{\mu\nu}F^{\mu\nu}
+\sum_{f=1}^{N_f }\Big(i\bar\psi_f\Dslash\psi_f-m_f\bar\psi_f\psi_f
\Big)\Big\}
}
and the covariant derivatives are appropriate to the representation $R_f$.

The problem before us is to compute
a Wilsonian effective action for the gauge theory on 
$ S^3\times S^1$ to the one loop order. We denote the length of $S^1$ by $L$
and the radius of $S^3$ by $R$.
The only zero modes in the
set-up belong to $A_0$, the gauge field component around
$ S^1$:
\EQ{
\alpha=\frac1{\text{Vol}\, S^3\times S^1}\int_{ S^3\times S^1} A_0\ .
} 
We can use a global gauge transformation to diagonalize $\alpha$:
\EQ{
\alpha=L^{-1}\text{diag}(\theta_i)\ .
\label{kii}
}
The $\theta_i$ are angular variables since there are large gauge
transformations (but periodic around $ S^1$)
that take $\theta_i\to\theta_i+2\pi$. Physically, the gauge
invariant quantities are built out of the Wilson loop 
$P=e^{i L \alpha}$ (the Polyakov loop in the case of thermal boundary
conditions) evaluated in the fundamental
representation:\footnote{Unless otherwise specified, traces are taken
  in the fundamental representation.}
\EQ{
\Tr\,P^n=\sum_{j=1}^Ne^{in\theta_j}\ ,
\label{wl}
}
On top of this there are 
additional large gauge transformations that are only periodic on
$S^1$ up
to a subgroup of the centre $\Z_N$ 
depending on the matter content of the theory. For adjoint matter,
this subgroup is the whole of $\Z_N$ and so
non-periodic 
large gauge transformations take $\theta_i\to\theta_i+2\pi/N$ and so 
transform $\Tr P$ by an $N$-th root of unity. Hence,
strictly speaking, the gauge invariant observables are, for example,
$|\Tr P|$.

The radiative corrections at the one loop level are obtained by
taking the constant mode \eqref{kii} as a background VEV and
integrating out all the massive modes of the fields. To this end, we
shift $A_0\to A_0+\alpha$ and then the one-loop
contribution involves the logarithm of the resulting functional
determinants which depend on $\alpha$ in a non-trivial way. 
As usual we have to fix the gauge in some way and we prefer to use a
different and more conventional approach
than that of \cite{Aharony:2003sx}. To this end we impose Feynman
gauge by adding the standard gauge fixing term 
\EQ{
S_\text{gf}={1\over g^2_{YM}}\frac1{2}\int
d^4x\sqrt{g}\,\text{Tr}\Big(\nabla_iA^i+\tilde 
D_0A_0\Big)^2,\
\label{gfa}
}
and appropriate ghosts. To the one loop level, we expand the action to
quadratic order in fluctuations. The gauge field part of the action,
including the ghosts, is 
\SP{
S_{\rm gauge}&=\frac1{g^2_{YM}}\int
d^4x\sqrt{g}\,\text{Tr}\,\Big[\tfrac12
A_0(-\tilde D_0^2-\Delta^{(s)})A_0\\ &
+\tfrac12A_i(-\tilde D_0^2-\Delta^{(v)})A^i+
+\bar c(-\tilde D_0^2-\Delta^{(s)})c\Big]\ .
\label{asd}
}
Here, $\Delta^{(s)}$ and $\Delta^{(v)}$ are the Laplacians
on $S^3$ for
scalar and gauge fields, respectively. The scalar Laplacian is simply
$\Delta^{(s)}=g^{-1/2}\partial_\mu(g^{1/2}\partial^\mu)$ whilst the vector
Laplacian is
\EQ{
\Delta^{(v)}A^i=\nabla_j\nabla^jA^i-R^i{}_jA^j\ ,
}
where $R_{ij}$ is the Ricci tensor of $S^3$. In the
above, $\tilde D_0=\partial_0+i\alpha$ and so includes the coupling to
the VEV

Each fluctuating field is expanded in terms of appropriate harmonics
on $ S^3\times S^1$ and a typical contribution to the effective action
is of the form
\EQ{
\pm\text{Tr}_R \log(-\tilde D_0^2-\Delta)\ ,
\label{sos}
}
the $\pm1$ being for $c$-number and Grassmann fluctuations, 
respectively and $\Delta$ is the Laplacian on
$ S^3$ appropriate to the tensorial nature of the field on $ S^3$;
either $\Delta^{(s)}$, $\Delta^{(v)}$ or $\Delta^{(f)}$. 
The background VEV $\alpha$ acts as a generator of the Lie
algebra of $SU(N)$ in the representation $R$ of the gauge group
appropriate to the field and the trace includes a trace over that
representation of the gauge group.
The eigenvalues of $\partial_0$ are
simply $2\pi i n/L$, $n\in\Z$, while the eigenvectors of the
Laplacian on $ S^3$ are labelled by the angular momentum $\ell$:
\EQ{
\Delta \psi_\ell=-\varepsilon_\ell^2\psi_\ell\ ,
}
and we denote their degeneracy as $d_\ell$. The
$\varepsilon_\ell$ and $d_\ell$ depend on the 
field type. We review the spectra of the appropriate Laplacians on an
arbitrary sphere in Appendix A.
For us, the relevant fields are scalars (more precisely
minimally coupled scalars), vectors and spinors and below we list the
relevant data:

(i) {\bf Scalars}. For minimally coupled scalars
$\varepsilon_\ell=R^{-1}\sqrt{\ell(\ell+2)}$ and the 
degeneracy $d_\ell=(\ell+1)^2$ with $\ell\geq0$. 

(ii) {\bf Spinors}. For the irreducible 2-component real
spinors,\footnote{A Majorana spinor on $ S^3\times  S^1$ corresponds to 2 such
  spinors on $ S^3$.} we have
$\varepsilon_\ell=R^{-1}(\ell+1/2)$ and
$d_\ell=\ell(\ell+1)$ with $\ell>0$. 

(iii) {\bf Vectors}. Here the situation is more complicated. 
A vector field $V_i$ can be decomposed
into the image and the kernel of the covariant derivative:
$V_i=\nabla_i\chi+B_i$, with $\nabla^iB_i=0$. The eigenvectors 
for the ``transverse part'', $B_i$, have
$\varepsilon_\ell=R^{-1}(\ell+1)$ and $d_\ell=2\ell(\ell+2)$ with
$\ell>0$. On the
other hand, the ``longitudinal part'' $\nabla_i\chi$ has
$\varepsilon_\ell=R^{-1}\sqrt{\ell(\ell+2)}$ with degeneracy
$d_\ell=(\ell+1)^2$ but with $\ell>0$ only.  

\begin{table}
\begin{center}\begin{tabular}{ccccc}\hline\hline
field & angular mom. & energy & degeneracy \\
\hline
$B_i $ & $ \ell>0 $ & $ (\ell+1)/R$ & $
2\ell(\ell+2) $ \\ 
$C_i $ & $ \ell>0 $ & $ \sqrt{\ell(\ell+2)}/R$ & $
(\ell+1)^2 $\\ 
$\bar c,c $ & $ \ell\geq0 $ & $ \sqrt{\ell(\ell+2)}/R$ & $
-2(\ell+1)^2 $\\ 
$A_0$ & $ \ell\geq0 $ & $ \sqrt{\ell(\ell+2)}/R$ & $
(\ell+1)^2 $\\
$\psi_\alpha $ & $ \ell>0 $ & $
\sqrt{(\ell+\tfrac12)^2+m^2R^2}/R$ & $ -2\ell(\ell+1) $\\
\hline\hline
\end{tabular}\end{center}
\caption{\small The fields, their angular momenta, energy and degeneracy
(with $\pm$ sign for $c$-number and Grassmann fluctuations)  
in the effective action. The fermion result is for a massive Majorana fermion on $S^3\times S^1$.}
\end{table}

It is a standard calculation using the identity
$\prod_{n=1}^\infty(1+x^2/n^2)=\sinh(\pi x)/(\pi x)$
to show that \eqref{sos} is equal, up to an infinite additive constant, to
\EQ{
\sum_{\ell=0}^\infty d_\ell\Big\{L \varepsilon_\ell
-2 \sum_{n=1}^\infty\frac1n 
e^{-n L \varepsilon_\ell}\Tr_\text{R}\big(P^n\big)\Big\}\ .
\label{saa}
}
The first term here involves the Casimir energy and since it is independent
of $\alpha$ will play no r\^ole in our story and we will subsequently
drop it. 

Notice that the $\ell>0$ contributions from $A_0$, $C_i$ and the
ghosts all cancel leaving only a net contribution from the $\ell=0$
modes of the
form\footnote{This part is precisely the exponentiation of the Jacobian that
converts the integrals over the $\theta_i$ into an integral over the
unitary matrix $P=\text{diag}(e^{i\theta_i})$:
\EQ{
\int\prod_{i=1}^Nd\theta_i\,\exp\Big\{
\sum_{n=1}^\infty\frac1n\Tr_\text{adj}\big(P^n\big)\Big\}
\propto\int\prod_{i=1}^Nd\theta_i\,\prod_{i<j}\sin^2\Big(
\frac{\theta_i-\theta_j}2\Big)
=\int dP\ .
}
However, we will leave the Jacobian in the exponent since it must be
considered as part of the effective action for the eigenvalues.}
\EQ{
\sum_{n=1}^\infty\frac1n
\text{Tr}_\text{adj}\big(P^n\big)\ .
\label{jac}
}
The remaining gauge modes are the vector modes $B_i$ and 
the fermions. Using the 
sum \eqref{saa} and including the Jacobian term
in \eqref{jac}, the full effective action is then 
\EQ{
S(\alpha)=\sum_{n=1}^\infty\frac1n\Big\{
\big(1-z_v(n L/R)\big)\Tr_\text{adj}\big(P^n\big)
+\sum_{f=1}^{N_f }
z_f(n L/R,m_fR)\Tr_{R_f}\big(P^n\big)\Big\}\ .
\label{ioi}
}
In the above, we have defined
\EQ{
z_v(L/R)=
2\sum_{\ell=1}^\infty\ell(\ell+2)e^{-L(\ell+1)/R}=
\frac{6e^{-2 L/R}-2e^{-3 L/R}}{(1-e^{- L/R})^3}\ ,
\label{jhh}
}
and for the spinors
\EQ{
z_f( L/R,mR)=2\sum_{\ell=1}^\infty\ell(\ell+1)e^{-L 
\sqrt{(\ell+1/2)^2+m^2R^2}/R}\ .
}
We have assumed that the fermions have periodic boundary conditions
around $S^1$. If one wanted to describe the case of finite temperature the
fermions have anti-periodic boundary conditions and
$z_f(n L/R,m_fR)$ must be multiplied by an additional factor of $(-1)^n$. 

Notice for the vector modes, we are able to perform the sum over the
angular momentum, but for the fermion modes this is not
possible due to the non-zero mass. 
However, a useful expression for the fermionic contribution
can be obtained by applying a version of the Abel-Plana formula which
is proved in Appendix B appropriated to a function with branch points
on the imaginary axis:
\EQ{
\sum_{\ell=0}^{\infty} f(\ell+\tfrac12) = \int_0^{\infty} dx\,
f(x) - i \int_0^{\infty}dx \, \frac{f(i x +
  \varepsilon) - f(-i x - \varepsilon)}{e^{2 \pi x} + 1}\ . 
\label{abelp}
}
In the above $\varepsilon$ is positive, real, and infinitesimal. Applying
this formula to the function
\EQ{
f(\ell)=2\ell(\ell+1/2)e^{-L \sqrt{(\ell+1/2)^2+m^2R^2}/R}\ ,
}
gives the integral representation
\SP{
&z_f(L/R,mR)\\
&=2\int_0^\infty dx\,(x^2-\tfrac14)e^{-L \sqrt{x^2+m^2R^2}/R}
+4\int_{mR}^\infty dx\frac{x^2+\tfrac14}{e^{2\pi
    x}+1}\sin(L \sqrt{x^2-m^2R^2}/R)\\
&=\frac{2m^2R^3}{L}K_2(L m)-\frac{mR}2K_1(L m)+
4\int_{mR}^\infty dx\frac{x^2+\tfrac14}{e^{2\pi
    x}+1}\sin(L \sqrt{x^2-m^2R^2}/R)\ .
\label{fex}
}
Note that in our case the function $f(ix)$ is real for $x<mR$ and 
becomes imaginary
for $x>mR$ and so the lower limit of the second 
integral has been taken to be $mR$. 

There are two consistency 
checks we can make on the integral expression \eqref{fex}. Firstly, in
the massless limit, $m\to0$, we have
\EQ{
z_f(L /R,0)=\frac{4e^{-3 L/2R}}{(1-e^{-L/R})^3}
\equiv\sum_{\ell=1}^\infty 2\ell(\ell+1)e^{-L(\ell+1/2)/R}\ .
}
Secondly in the limit $R\to\infty$ with fixed $m$ and $L$, we have
\EQ{
z_f(L/R,mR)\longrightarrow \frac{2m^2R^3}{L}K_2(L m)\ ,
}
which is the expression that one obtains by working directly on
$\R^3\times S^1$ \cite{Myers:2009df}. 

\section{Phase Structure of QCD(Adj)}

In this section, we apply our result \eqref{ioi} to the particular case of a
theory with $N_f $ adjoint fermions with equal masses. In this case, we
have

\EQ{
\Tr_\text{adj}\big(P^n\big)=\sum_{ij=1}^N\cos(n(\theta_i-\theta_j))
}
and so
\EQ{
S(\theta_i)=\sum_{n=1}^\infty\frac1n
\Big(1-z_v(n L/R)+N_f 
z_f(n L/R,m_fR)\Big)\sum_{ij=1}^N\cos(n(\theta_i-\theta_j))\ .
\label{ioi2}
}
The phase structure is determined by minimizing this 
with respect to the $\{\theta_i\}$. Since $N$ is large, it is more
appropriate to describe the configuration in terms of a density
$\rho(\theta)$ normalized so that
\EQ{
\int_0^{2\pi}d\theta\,\rho(\theta)=1\ .
}
In this case, we can write the effective action as  
\EQ{
S[\rho(\theta)]=N^2\int d\theta\,\int
d\theta'\,\rho(\theta)\rho(\theta')\sum_{n=1}^\infty
\frac{f(n L/R,mR)}n\cos(n
(\theta-\theta'))
\label{act}
}
In the above, we have defined the function
\EQ{
f(L/R,mR)=1-z_B(L/R)+N_f z_F(L/R,mR)\ .
}

It is useful to Fourier analyze the density:
\EQ{
\rho(\theta)=\frac1{2\pi}\,\sum_{n=-\infty}^\infty\rho_ne^{in\theta}\ ,
}
with $\rho_0 = 1$ and $\rho_n^*=\rho_{-n}$ for reality. The action then becomes
\EQ{
S[\rho(\theta)]=
\frac{N^2}2\sum_{n=1}^\infty \frac{f(n L/R,mR)}n|\rho_n|^2
}

The phase structure hinges on properties of the function
$f(L/R,mR)$ and specifically on its sign.
For large $L/R$ and any $m$, both $z_B,z_F\to0$ and so 
$f(L/R,mR)\to1$. The behaviour in the limit of small $L/R$
depends on how $m$ is scaled. If we keep $m L$ fixed then 
\EQ{
f(L/R,mR)\longrightarrow \frac{2m^2R^3}{L}\Big(N_f K_2(m L)
-\frac2{(m L)^2}\Big)\ .
\label{qse}
}
In this limit, $f$ is positive (negative) for $m L<a$ ($m L>a$), where
$a$ is the solution of  
\EQ{
N_f a^2K_2(a)=2\ ,
}
which is only possible if $N_f >1$. For $N_f =1$, $f$ is negative for all
$m L$ (in the limit of small $L/R$). If we consider small
$L/R$ but keep $mR$ fixed then
\EQ{
f(L/R,mR)\longrightarrow\frac{4R^3}{L^3}(N_f -1)-
\frac{m^2R^3N_f }L+\cdots\ .
}
Note that this is positive for $N_f >1$.

\begin{figure}[ht] 
\centerline{\includegraphics[width=2.5in]{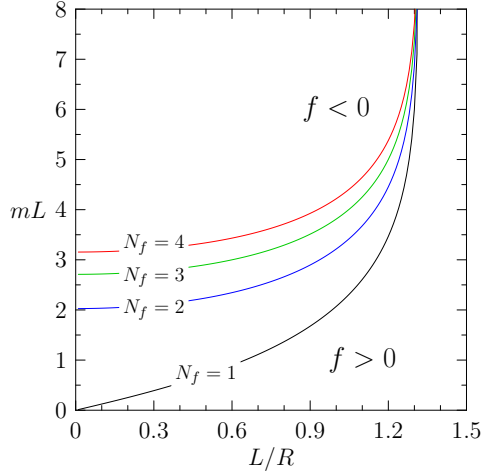}}
\caption{\small The lines where $f=0$ indicating the regions
  where $f>0$ and $f<0$ in the 
$(L/R,m L)$ plane for $N_f =1,\ldots,4$.}\label{fig4}
\end{figure}

It will be useful to chart the 
phase diagram initially in the 
$(L/R,m L)$ plane and in Figure \ref{fig4} we show the
corresponding regions for which $f$ is positive and negative, for
$N_f =1,2,3,4$. 

Having charted the region where $f$ is positive and negative we can
now build up a picture of the phase
structure. It will be useful to define $f_n\equiv f(n L/R,mR)$ so that
\EQ{
S=\frac{N^2}2\sum_{n=1}^\infty\frac{f_n}n|\rho_n|^2\ .
}
In the region where all the $f_n>0$, $n=1,2,\ldots$, it is clear that
the action will be minimized when all the Fourier modes vanish $\rho_n=0$, except $\rho_0 = 1$, {\it
  i.e.\/}~$\rho(\theta)=1/2\pi$. This phase is the weak-coupling
manifestation of the confining phase where the centre symmetry is
unbroken and $\Tr P^n=0$, $n>0$. In view of this 
we will call it the ``confining phase''.

The confining phase covers the region of the phase diagram where all
the $f_p>0$. This is separated from the rest of the phase diagram by
the union of arcs where a given $f_p$ vanishes. The arc where $f_p=0$
extends between 2 multi-critical points where
$f_p=f_{p+1}=0$ and $f_{p-1}=f_p=0$, respectively. 
As one crosses the contour on which $f_p=0$ a
phase transition occurs where the density
$\rho(\theta)$ develops $p$ gaps, that is $p$ intervals around the
circle on which $\rho(\theta)$ vanishes. We call the resulting phase
the $\Z_p$ phase since at strong coupling it would be identified with
a partially confined phase where the $SU(N)$ gauge group is 
broken to a subgroup $SU(p)$ that confines and a 
$\Z_p$ subgroup of the centre symmetry remains unbroken. The signature
of the $\Z_p$ phase is that the order parameters behave as
\SP{
&\Tr\, P^n=0\ ,~~~~~~n/p\not\in\Z\ ,\\
&\Tr\,P^n\neq 0\ ,~~~~~~n/p\in\Z\ .
}

The detailed argument of why such a transition
occurs follows as a generalization of the transition from the uniform
to one gap phase described in 
\cite{Aharony:2003sx} and we include it for completeness. The important point
is that the configuration space $\{\rho_n\}$ has a non-trivial
boundary which encloses the allowed region surrounding the
origin because the density $\rho(\theta)$ cannot be
negative. The boundary region
consequently consists of distributions for which $\rho(\theta)$
vanishes at a subset of points (including finite intervals) around the
circle. Furthermore, it is clear that allowed region in the
configuration space is a convex region, since if $\rho_1(\theta)$ and
$\rho_2(\theta)$ are allowed then so is
$t\rho_1(\theta)+(1-t)\rho_2(\theta)$, $0\leq t\leq1$. 

\begin{figure}[ht] 
\centerline{\includegraphics[width=2.5in]{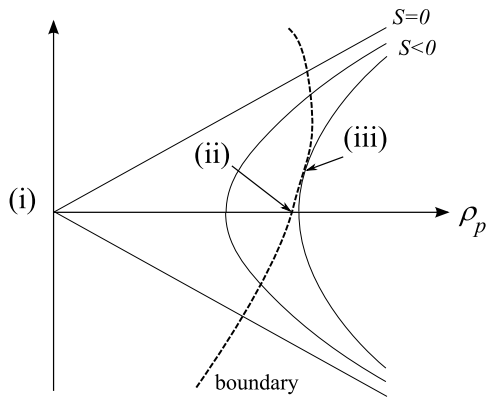}}
\caption{\small The structure of configuration space showing
  $|\rho_p|$ and one additional direction.
 The boundary is indicated by the dotted line and it is
  important that the allowed region is convex. For $f_p<0$, 
the lines of vanishing action define a cone
and the lines of constant negative action being hyperbolae therein. 
The density (i) is the uniform distribution characteristic of the
  confining phase; (ii) is the density with $|\rho_p|=\tfrac12$ 
which lies at the
  boundary of configuration space; and (iii) is the density with
  minimal action as lying at the boundary of configuration space where
  the lines of constant $S$ lie tangent to boundary.}\label{fig8}
\end{figure}

\begin{figure}[ht] 
\centerline{\includegraphics[width=5in]{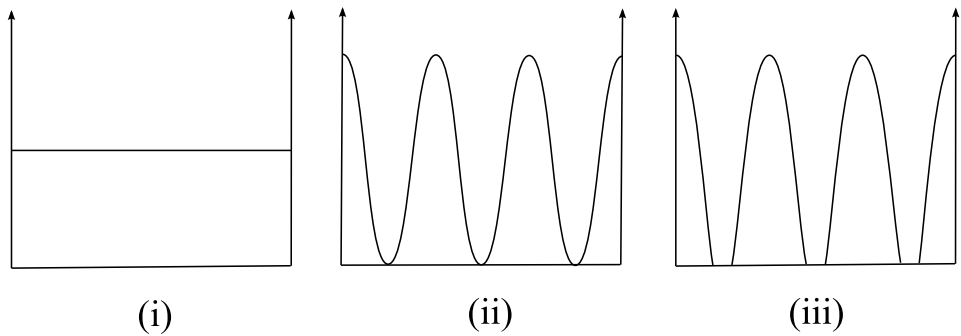}}
\caption{\small The behaviour of the density across the  
transition at $f_p=0$: (i) 
the uniform density in the confining phase and (iii) the
$\Z_p$ phase (for $p=3$). At the transition point (ii) 
the mode $\rho_p$ becomes
massless and the density develops $p$ zeros as shown in the 
middle.}\label{fig5}
\end{figure}

Let us consider the transition across the line where $f_p=0$. On the
confining phase side of the transition $f_p>0$ and
$\rho(\theta)=1/2\pi$. At the transition point, the action becomes
independent of the Fourier component $\rho_p$ and so the one complex parameter
family of densities
\EQ{
\rho(\theta)=\frac1{2\pi}\big(1+\rho_pe^{ip\theta}+\rho_p^*e^{-ip\theta}\big)
\label{sdd}
}
for $0\leq|\rho_p|\leq\tfrac12$ 
all have the same (vanishing) action. As $f_p$
becomes negative, then the points with $S=0$ in
the configuration space is a cone whose angle opens as 
$f_p$ becomes more negative. 
The locus of configurations with $S<0$ correspond to
hyperbolae lying inside this cone. It follows that the configuration
with minimal action lies on the boundary of the configuration space at
a point where one of the hyperbola lies tangent to the boundary,
{\it i.e.\/}~on a distribution where $\rho(\theta)$ vanishes at a subset
of points. This is illustrated in Figure \ref{fig8}. 
We conclude that new configuration is
continuously connected to the $|\rho_p|=\tfrac12$ 
density in \eqref{sdd} and so must
have precisely $p$ gaps. By
symmetry this phase will be invariant under a $\BZ_p$ subgroup of the
centre symmetry. A schematic view of the 
transition appears in Figure \ref{fig5}.
The transition is first order since the effective action is
discontinuous across the transition. In the confining phase 
we have $S=0$, whereas just above the transition to leading order 
it is sufficient to take the density to be \eqref{sdd} with
$|\rho_p|=\tfrac12$ and plug this into the action to get
\EQ{
S=\frac{N^2}{4p}f_p(L/R,mR)\ ,
}
because the density itself only changes at a higher order.
Since $f_p(L/R,mR)$ has non-vanishing first derivatives in
$\delta L$ and $\delta m^2$ the derivatives of $S$ change
discontinuously across the transition implying that it is first order.

As we remarked above the critical lines $f_p=0$ and $f_{p+1}=0$ cross at 
multi-critical points where $f_p=f_{p+1}=0$, and at these points 
the confining, $\Z_p$ and $\Z_{p+1}$ phases are all continuously connected via
the density 
\EQ{
\rho(\theta)=\frac1{2\pi}\big(1+\rho_pe^{ip\theta}+\rho_p^*
e^{-ip\theta}+\rho_{p+1}e^{i(p+1)\theta}+
\rho_{p+1}^*e^{-i(p+1)\theta}\big)\ .
\label{sdd2}
}
By continuity, 
it must be that there are lines of
first order phase transitions that separate the
partially confined phases
and which end on the critical points that lie somewhere between the 
continuation of the $f_p=0$ and $f_{p+1}=0$ lines.\footnote{The argument 
is as follows, in the vicinity of the
  critical point we have  $S=N^2f_p/4p$ and
  $N^2f_{p+1}/4(p+1)$, respectively, in the $\Z_p$ and $\Z_{p+1}$
  phases. A first order transition occurs when $f_p/p=f_{p+1}/(p+1)$
  which must necessarily be in a region where $f_p<0$ and $f_{p+1}<0$,
  in other words somewhere between the continuations of the $f_p=0$
  and $f_{p+1}=0$ lines.}
The actual positions of
these lines of transition will depend in detail on the gapped
distributions. However, in the limit $R\to\infty$ with $m$ and $L$
fixed, we know from \eqref{qse} that up to an overall factor
$f(L/R,mR)\propto g(m L)$. Hence, these lines of first order
transitions must asymptote to lines of constant $m L$.

In the $\Z_p$ phase, as 
$L/R\to0$ we expect the gaps will grow and in this limit
the density $\rho(\theta)$ in the $\Z_p$
phase will only have support at $p$ equally spaced points around
the circle
\EQ{
\rho(\theta)\longrightarrow \frac1p\sum_{j=0}^{p-1}\delta(\theta-2\pi
j/p)\ .
}
In this case, the line of transitions will occur for $m L$
being the solution of the equation
\EQ{
\sum_{j=1}^\infty \frac{f(j p L/R,mR)}{jp}=
\sum_{j=1}^\infty \frac{f(j(p+1)L/R,mR)}{j(p+1)}\ ,
}
or more concretely since $L/R\to0$ we can use \eqref{qse} to get the
conditions 
\SP{
\sum_{j=1}^\infty &
\frac1{(jp)^2}\Big(N_f K_2(p j m L)-\frac2{(p j m L)^2}
\Big)\\ &=\sum_{j=1}^\infty
\frac1{(j(p+1))^2}\Big(N_f K_2((p+1) j m L)-\frac2{((p+1)j m L)^2}
\Big)\ .
}

\begin{figure}[ht]
\centerline{(a)\includegraphics[width=2.5in]{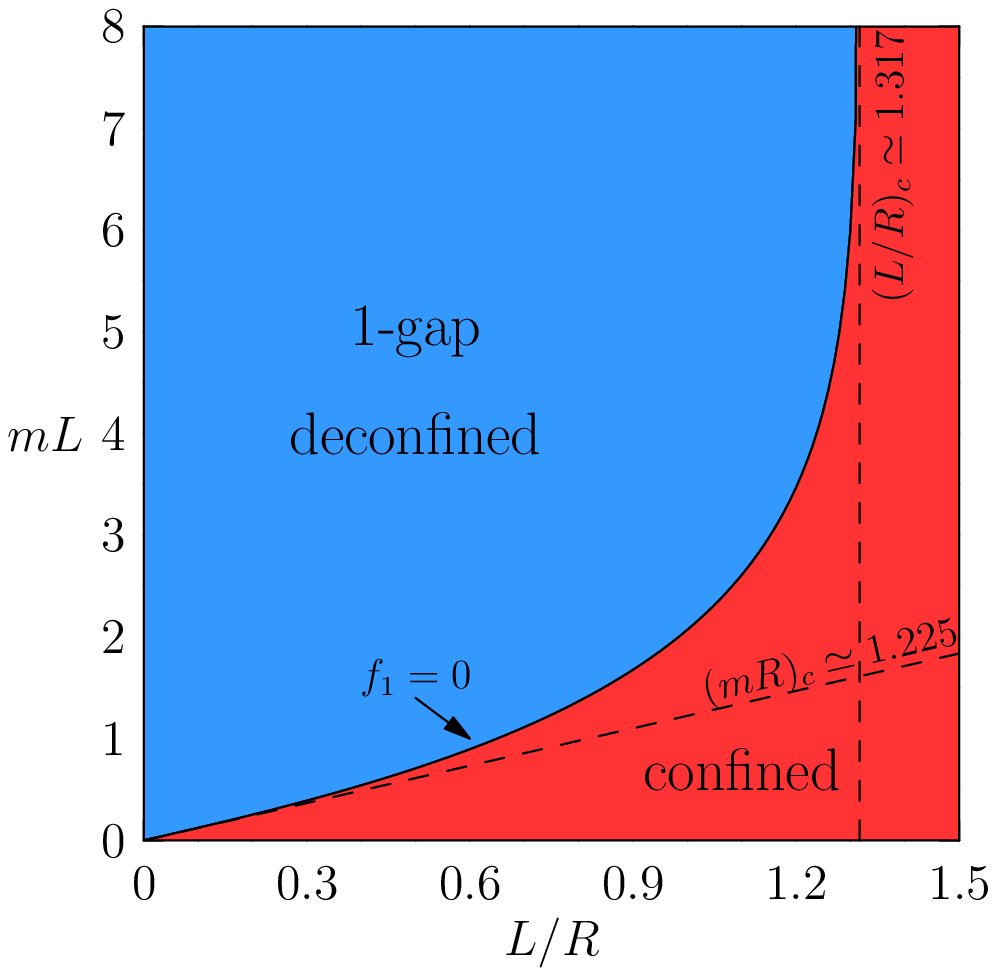}
\hspace{0.2cm}(b)\includegraphics[width=2.5in]{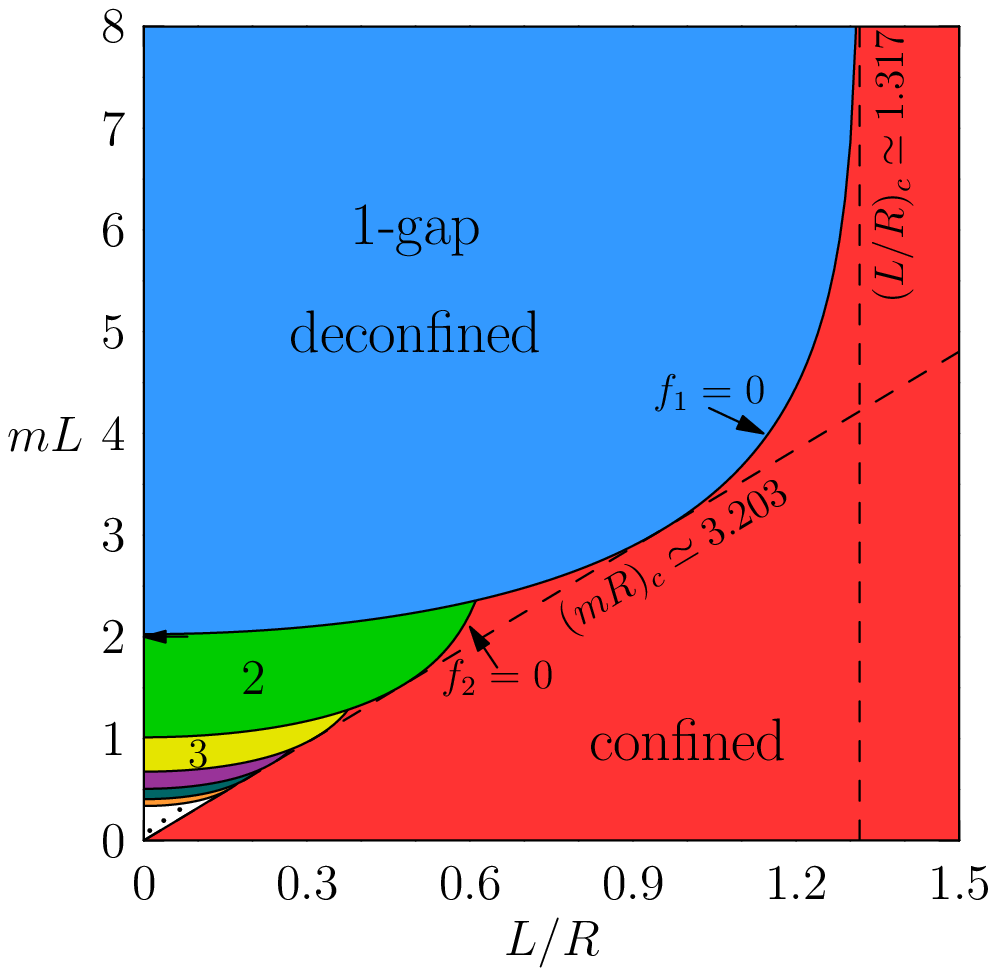}}
\caption{\small The phase diagrams in $(L/R,m L)$ coordinates
for (a) $N_f =1$ and (b) $N_f  = 2$. We have shown the transitions
between the $\Z_p$ and $\Z_{p+1}$ phases along the continuation of the
$f_p=0$ line for simplicity since this seems to be a good approximation
and matches the value calculated in the  
${\field R}^3 \times S^1$ from \cite{Myers:2009df} indicated by the 
arrow pointing to the $m L$-axis.}\label{fig3} 
\end{figure}

The phase diagram in the $(L/R,mL)$ plane is shown in Figure
\ref{fig3} for the two distinct cases $N_f =1$ and $N_f =2$. 
In the former case, only
the phase with one gap appears which at strong coupling is identified with
the de-confined phase since the centre symmetry is completely broken. 
For $N_f >1$ all the $\Z_p$ phases appear for
$p=1,2,\ldots$.

The transition between the confined phase and $k$-gap phases occurs
along the $f_k = 0$ curve. The line of transitions between $k$-gap and
$k+1$-gap phases only slightly differs from the $f_k = 0$ curves of
Figure \ref{fig3}. From the ${\field R}^3 \times S^1$ result in
\cite{Myers:2009df} the transition between the $1$ and $2$-gap phases
occurs for $m L \simeq 2.020$, where as the $m L$ asymptote of the
$f_1 = 0$ curve lies at $m L \simeq 2.027$. This difference is only
barely visible in Figure \ref{fig3}. The confined phase is always
favoured below a critical line with slope $(m R)_{c}$ which increases
with $N_f$. The $(L/R)_{c} \simeq 1.317$ line in both figures
indicates the value of the deconfinement temperature of the pure
$SU(\infty)$ Yang-Mills theory determined in \cite{Aharony:2003sx}. 

\begin{figure}[ht]
\centerline{(a)\includegraphics[width=2.5in]{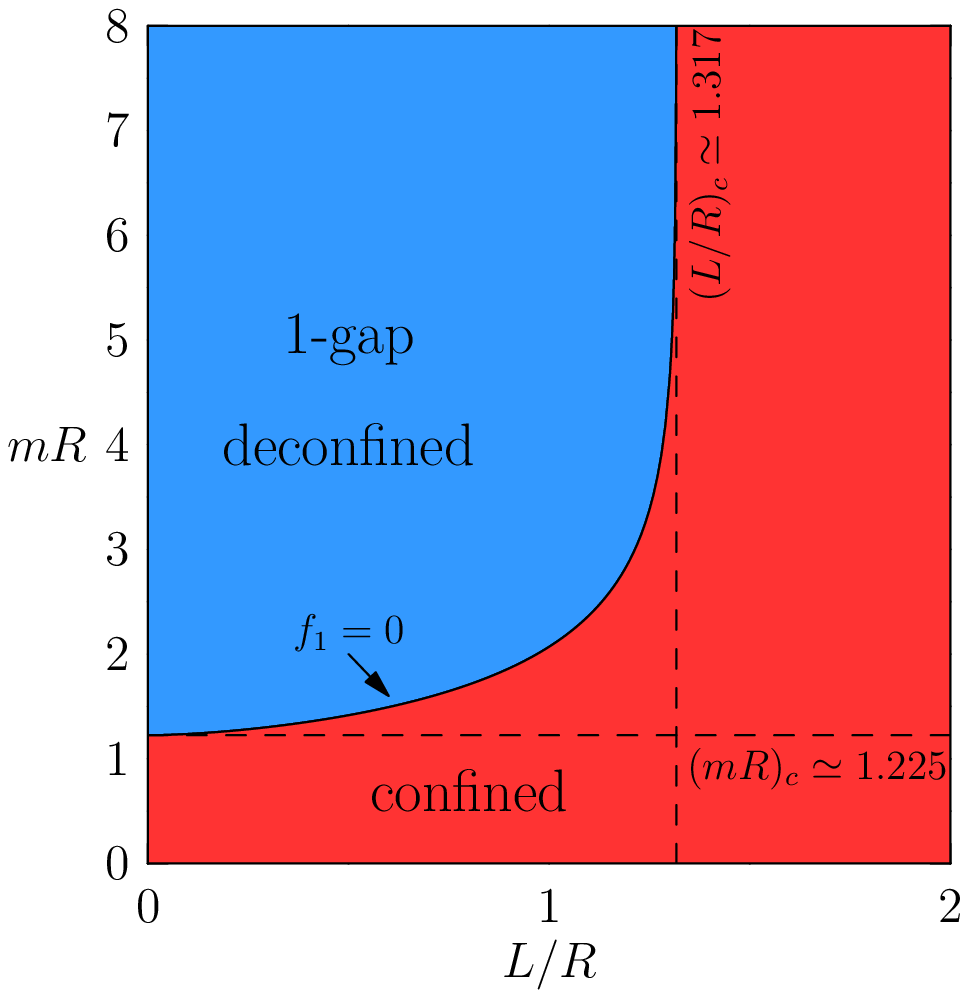}
\hspace{0.2cm}(b)\includegraphics[width=2.5in]{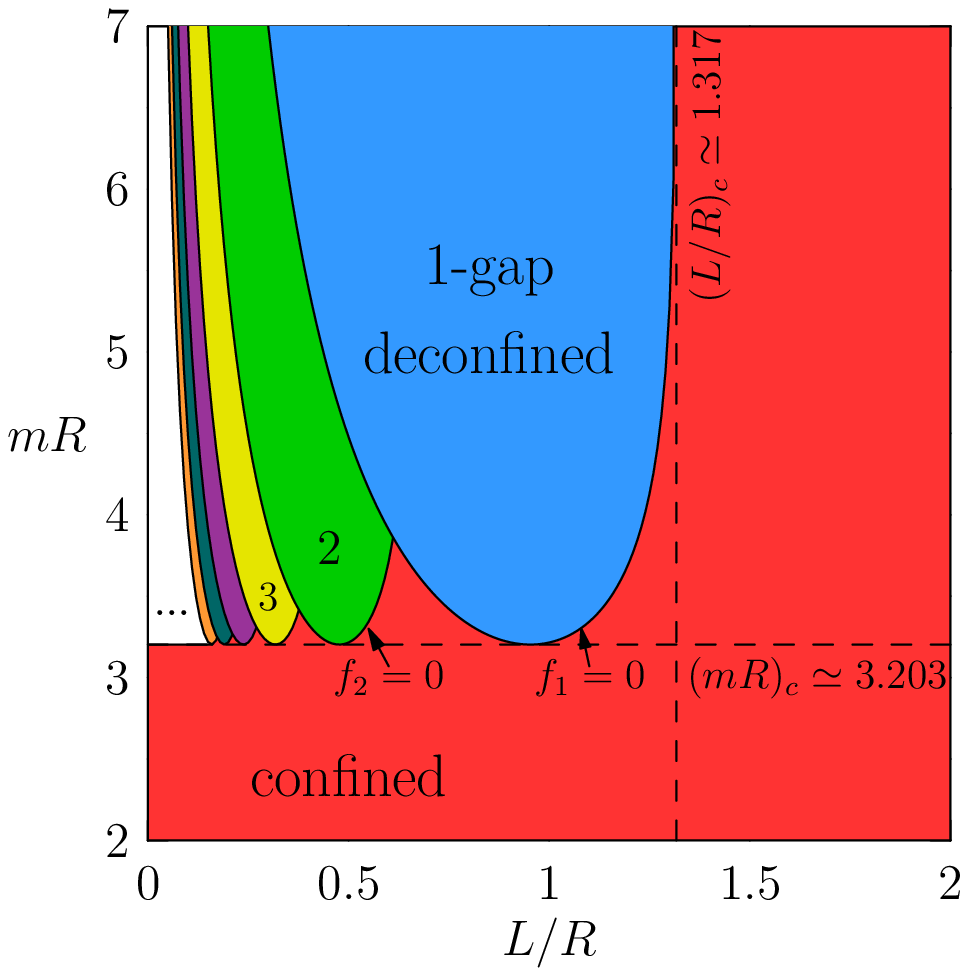}}
\caption{\small The phase diagram in $(L/R,m R)$ coordinates
for (a) $N_f =1$ and (b) $N_f  = 2$. As above, we have shown the transitions
between the $\Z_p$ and $\Z_{p+1}$ phases along the continuation of the
$f_p=0$ which seems to be a good approximation.}\label{fig1}
\end{figure}

 In Figure \ref{fig1}, we re-plot the same phase diagrams in the
$(L/R,mR)$ plane. Again it is clear that only the
confining phase exists for small enough $mR$. 
The basic form of the phase diagram can be understood intuitively as
follows. The important point is that periodic fermions contribute
positively to $f(L/R,mR)$ and consequently 
tend to act so as to disorder the Polyakov loop, counteracting the
effect of the gauge field, and 
favour the confined phase. However, their effect goes away as $m$
increases due to decoupling.
Consequently, for large fermion mass $m$, the matter fields decouple and we
recover the confinement/de-confinement of the pure gauge theory. Whereas
for small mass the fermions win in the
competition with the gauge fields and disorder the Polyakov loop. The
most striking result of our analysis is that the
confining phase extends all the way down to small $L/R$ as long as the
fermion mass in units of $1/R$ is below a critical value, $m\lesssim1.225/R$ for $N_f=1$,
$m\lesssim3.203/R$ for $N_f=2$. 


\section{Finite $\boldsymbol N$}

It is useful to also consider QCD(Adj)
at finite $N$ for several reasons:
(i) we can determine how the large $N$ limit is approached and develop some
intuition of when finite $N$ results start to approximate those in the
infinite $N$ limit, (ii) with finite $N$ and finite volume better
qualitative comparison with lattice results is possible, and (iii) it is
possible to compare with finite $N$ results on ${\field
  R}^3 \times S^1$ by considering the limit $R \rightarrow \infty$. 

To build intuition on how the infinite $N$ limit is approached it is
helpful to remove fermions for the moment and consider pure Yang-Mills
theory at finite $N$. In the strong coupling limit the de-confinement
phase transition has been observed in lattice simulations and it is
believed that the transition is second order for $N = 2$ and first
order for $N \ge 3$ \cite{Lucini:2005vg}.  

The weak-coupling analogue of the de-confinement transition of pure Yang-Mills theory is
observable from perturbation theory on small volume manifolds at weak coupling. In
\cite{Aharony:2003sx} the authors calculated the de-confinement
temperature from one-loop perturbation theory on $S^3 \times S^1$ and
found it to occur at $T_d R = 0.75932$ in the large $N$ limit. In \cite{Aharony:2005bq} the
same authors computed higher loop corrections 
to show that the large $N$ 
de-confinement transition is first order in the weak coupling limit.

\begin{figure}[t]
  \hfill
  \begin{minipage}[t]{.45\textwidth}
    \begin{center}  
      \includegraphics[width=0.95\textwidth]{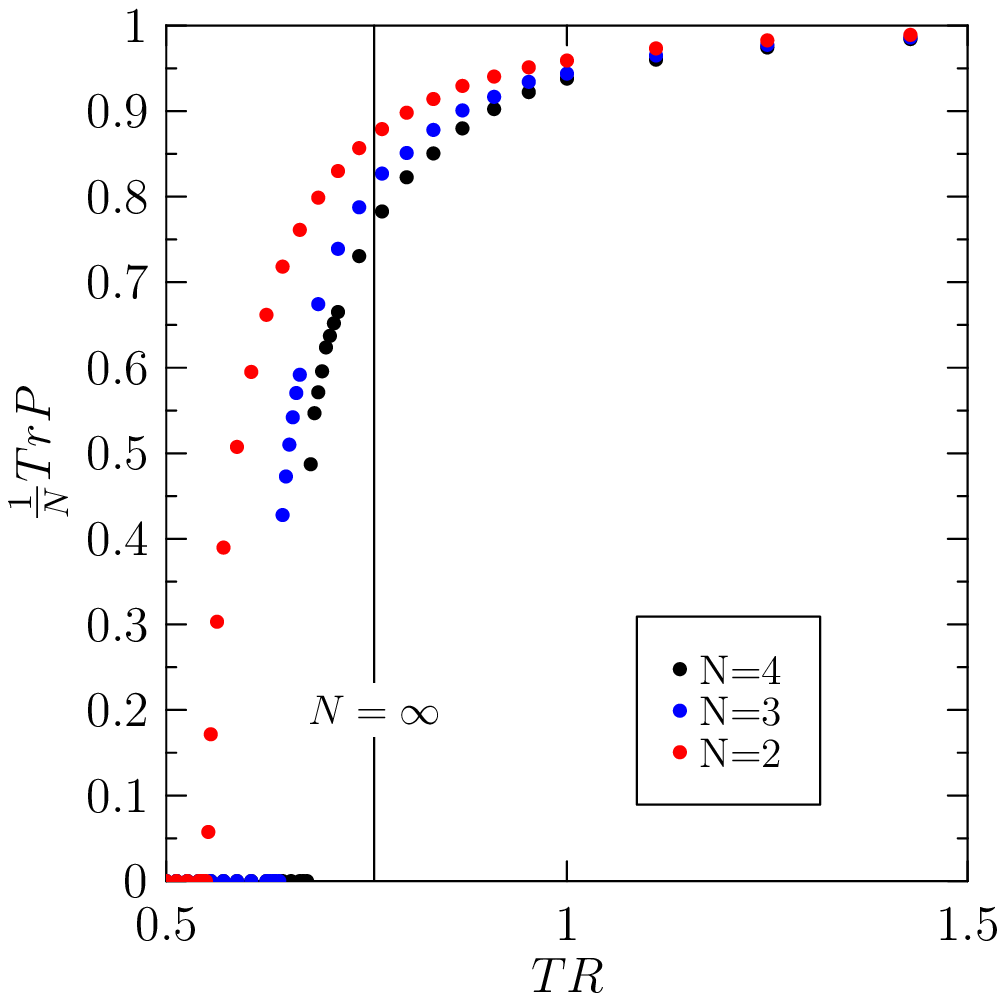}
    \end{center}
  \end{minipage}
  \hfill
  \begin{minipage}[t]{.45\textwidth}
    \begin{center}
\includegraphics[width=0.95\textwidth]{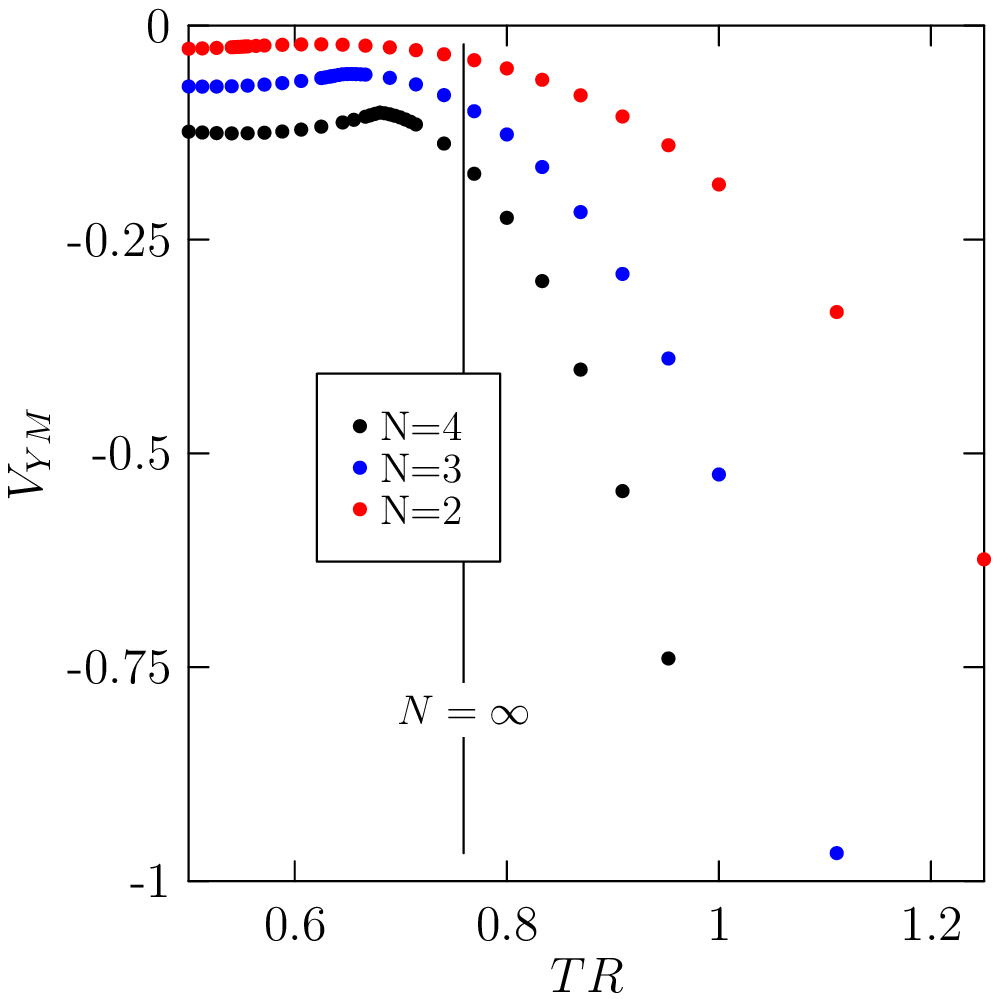}
    \end{center}
  \end{minipage}
  \hfill
  \caption{\small Yang-Mills theory for $N = 2, 3, 4$: (Left) ($T R$,
    $\Tr P$). For $N = 2$: $0.549 < T_d R < 0.552$, $N = 3$: $0.641 <
    T_d R < 0.645$, $N = 4$: $0.676 < T_d R < 0.680$; (Right) ($T R$,
    $V_{YM}$): $V_{YM}$ is the free energy density minus the ${\rm const}/ T R$ Casimir term. For larger values of $N$ this result appears increasingly compatible with a first order transition.} 
  \label{yang-mills}
\end{figure}

By numerically minimizing the Yang-Mills effective potential with respect to the
Polyakov loop eigenvalues it is possible to compute the trace of the
Polyakov loop order parameter as a function of the temperature as
shown in Figure \ref{yang-mills} (Left) for $N = 2, 3, 4$ (See Appendix \ref{nmin} for a short discussion of numerical minimization of the effective potential for finite $N$). The
discontinuity in the trace of the Polyakov loop is a clear indication
of the de-confinement transition even for $N = 2$. Increasing $N$
causes $T_d R$ to approach the large $N$ result from
\cite{Aharony:2003sx}. Near to the transition the points are separated
by $\Delta L/R = 0.01$. With this resolution the $N = 2$ transition
appears much smoother than for $N = 3, 4$, as we might expect from
lattice results. However, it is not possible to decipher the order of
the transitions for certain without taking the infinite volume limit. 

The corresponding Yang-Mills effective potential (eq. \eqref{ioi2} with $N_f = 0$ or $m = \infty$, which
doesn't include the ${\rm const} / T R$ Casimir term) is shown in Figure \ref{yang-mills}
(Right). The slight hump (most visible in the $N = 4$ result)
indicates the approximate location of the de-confinement transition. It
appears to become more well-defined with increasing $N$. Even though
the finite $N$ transitions shown in figure \ref{yang-mills} are not
genuine phase transitions since this system does not have an infinite
number of degrees-of-freedom, true transitions can be obtained by
taking $N$ or $R$ to infinity. The sharpness of the hump in the
effective potential serves as an indicator of how well the finite $N$
transition approximates a true infinite $N$, or infinite volume
transition. 

\begin{figure}[t]
  \hfill
  \begin{minipage}[t]{.45\textwidth}
    \begin{center}  
      \includegraphics[width=0.95\textwidth]{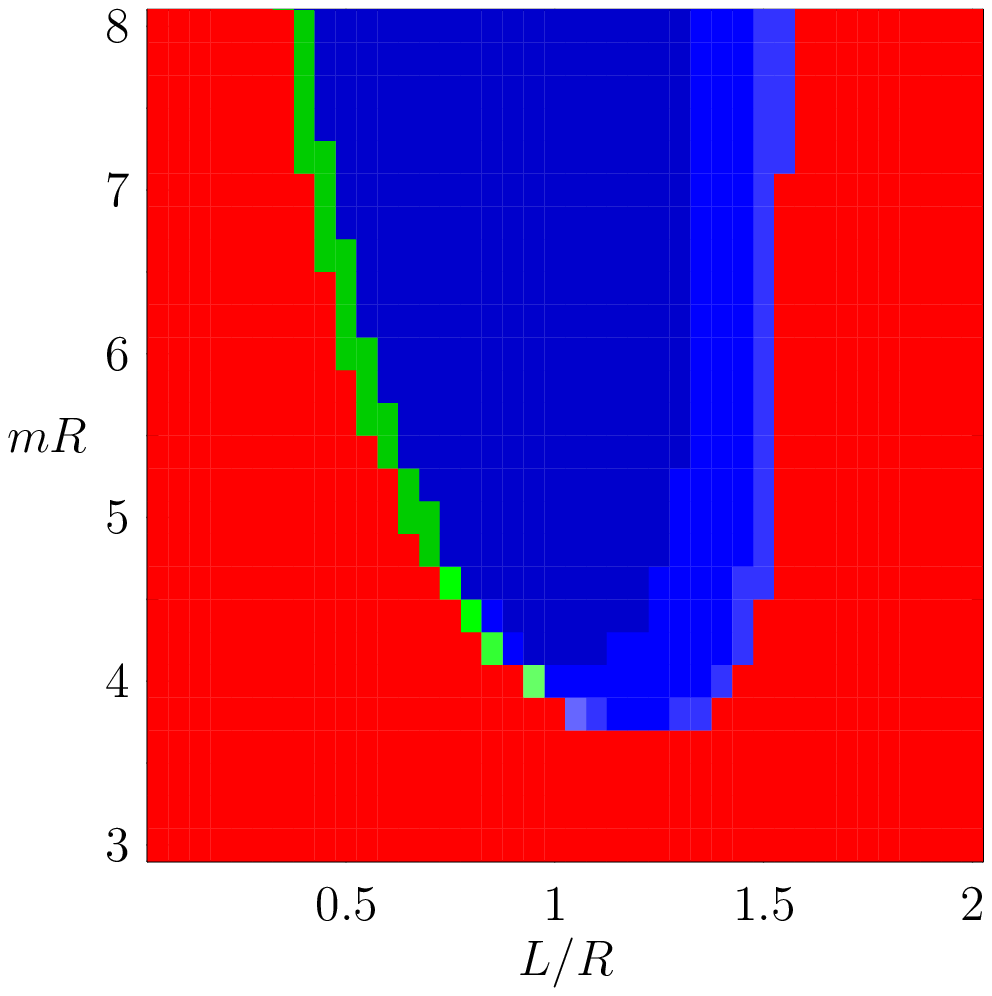}
    \end{center}
  \end{minipage}
  \hfill
  \begin{minipage}[t]{.45\textwidth}
    \begin{center}
\includegraphics[width=0.95\textwidth]{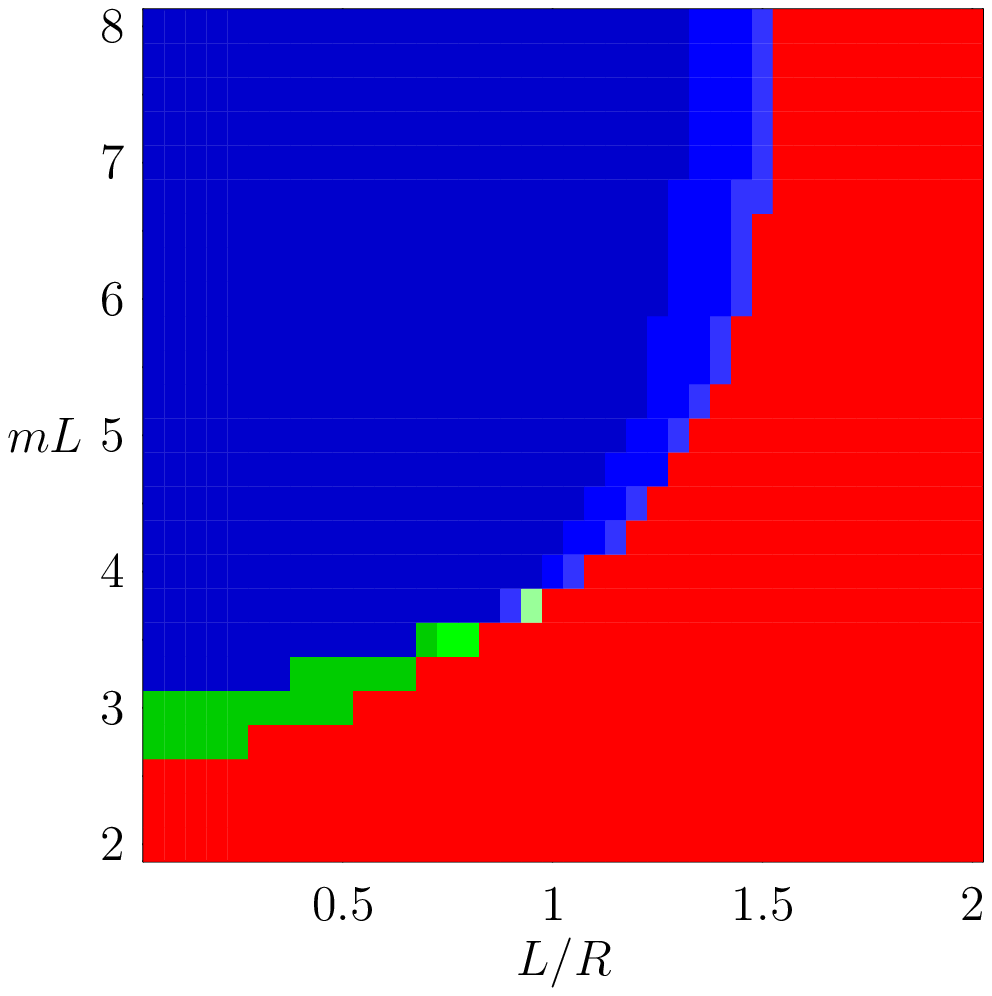}
    \end{center}
  \end{minipage}
  \hfill
  \caption{\small Phase diagram of QCD(Adj) for $N = 3$, $N_f  = 4$: (Left) in the ($L / R$, $m R$) plane with $3.6 < (m R)_c < 3.8$; (Right) in the ($L / R$, $m L$) plane with $2.5 < (m L)_c < 2.75$.}
  \label{adjn3nf4}
\end{figure}

It is important to emphasize that the use of the saddle point approximation to determine the preferred configuration of the gauge field is not strictly valid in the limit of finite $N$. However, one can show that it is still a reasonable approximation for purpose of obtaining the phase diagram, even for $N = 2, 3$, by plotting the effective action in the complete configuration space of the $\theta_i$. The effective action has clear minima in the configuration space corresponding to the eigenvalues obtained with the saddle point approximation. The competition with other configurations is minimal. For $L/R$ below $(L/R)_c$ there is a clear distinction between the confined and gapped phases which can be shown by performing the integrals over the gauge fields to obtain the partition function, $Z$, then plotting $e^{-S} \Tr P / Z$ in the full configuration space and considering a small radius around the values of the $\theta_i$ corresponding to the minima of the effective action. Above $(L/R)_c$ the fluctuations of the Polyakov loop in the full configuration space of the $\theta_i$ are more severe, but taking the average in a small radius around the configurations determined by the saddle point approximation still results in $\langle \Tr P \rangle = 0$. The fact that the integrals over the gauge field can be solved numerically serves as a check of the results of the saddle point approximation. However, expectation values of the Polyakov loop, for example, will always be zero, so it is important to find the eigenvalues from the saddle point approximation, and compare with plots of $e^{-S} \Tr P / \int [d\theta] e^{-S}$, as a function of the $\theta_i$, such that the different $Z(N)$ phases can be distinguished. Using the saddle point approximation serves as a means of obtaining sharper phase transitions than would be observed by performing the full integrals over the gauge field configurations. It additionally picks out eigenvalues for a single minimum of the effective potential, avoiding the issue of finding a suitable order parameter for distinguishing the $Z(N)$ vacua (or the relevant subgroup) which would otherwise be necessary since averaging over the full configuration space gives $\langle \Tr P \rangle = 0$ in all phases.

For QCD(Adj) at finite $N$ we can perform an analysis similar to
what was done in the case of large $N$. The effective potential has
the exact same form except that we can't solve the path integral using
the saddle point approximation unless $N$ is large enough. Since we don't know when this is true we again use numierical minimization.

\begin{figure}[t]
  \hfill
  \begin{minipage}[t]{.45\textwidth}
    \begin{center}  
      \includegraphics[width=0.95\textwidth]{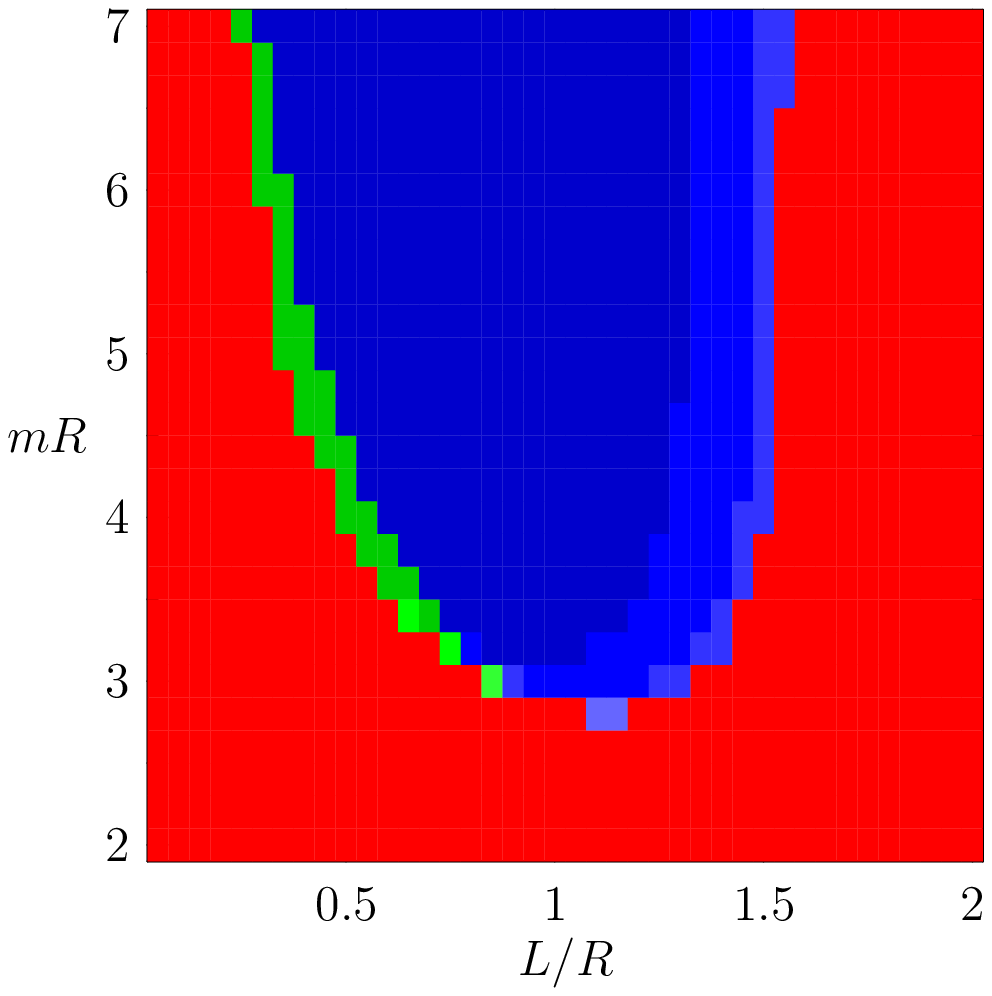}
    \end{center}
  \end{minipage}
  \hfill
  \begin{minipage}[t]{.45\textwidth}
    \begin{center}
\includegraphics[width=0.95\textwidth]{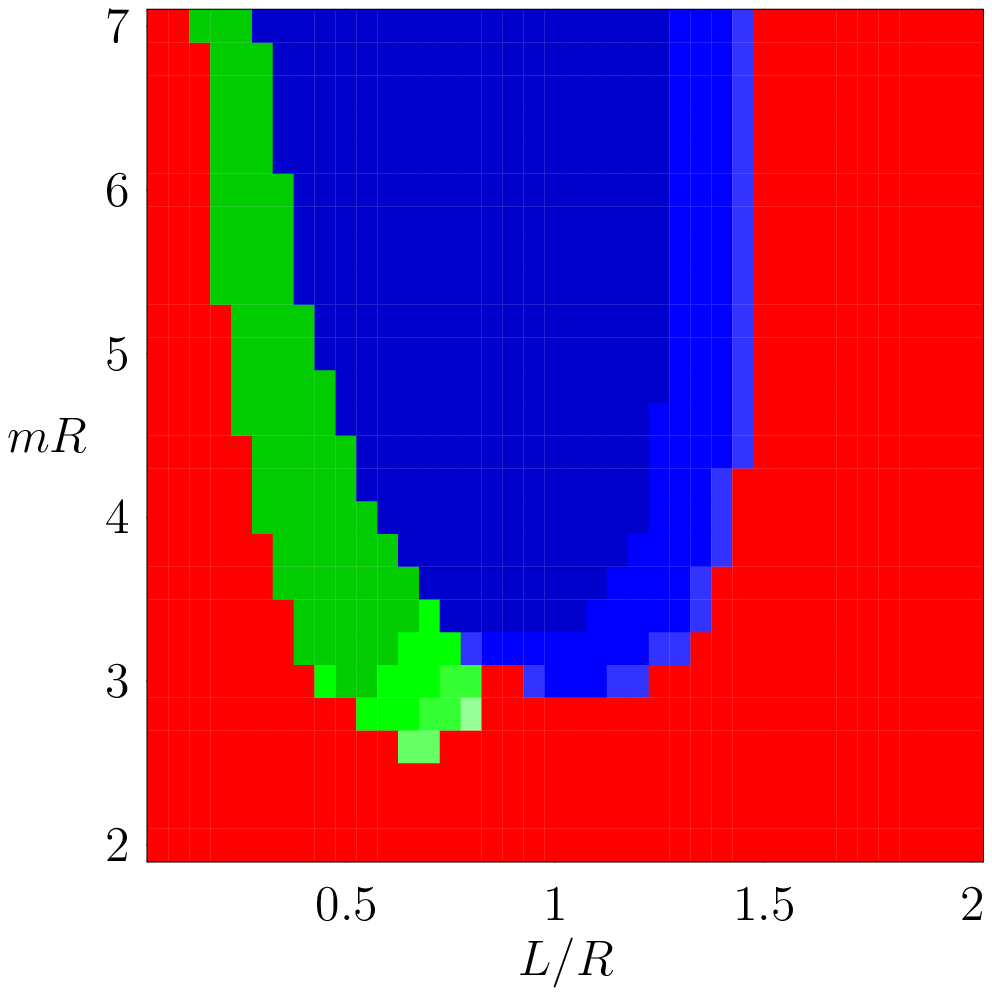}
    \end{center}
  \end{minipage}
  \hfill

  \hfill
  \begin{minipage}[t]{.45\textwidth}
    \begin{center}  
      \includegraphics[width=0.95\textwidth]{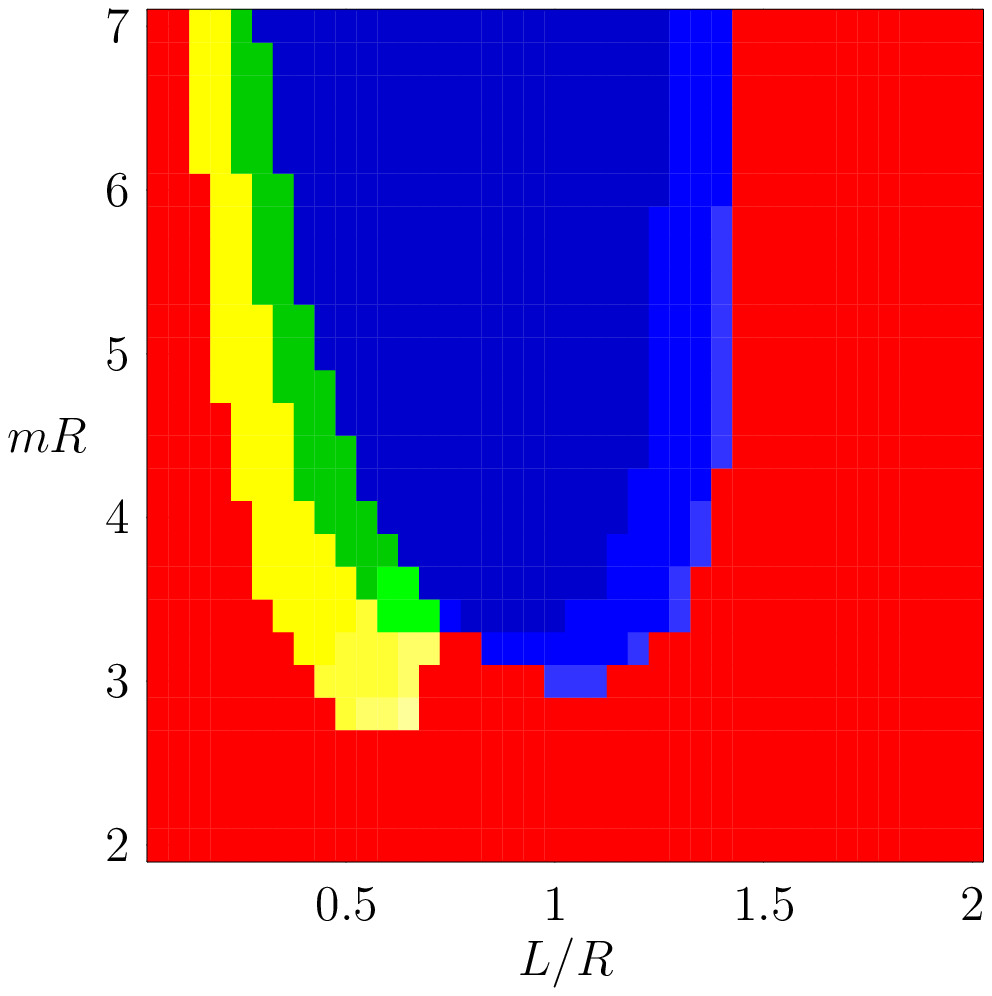}
    \end{center}
  \end{minipage}
  \hfill
  \begin{minipage}[t]{.45\textwidth}
    \begin{center}
\includegraphics[width=0.95\textwidth]{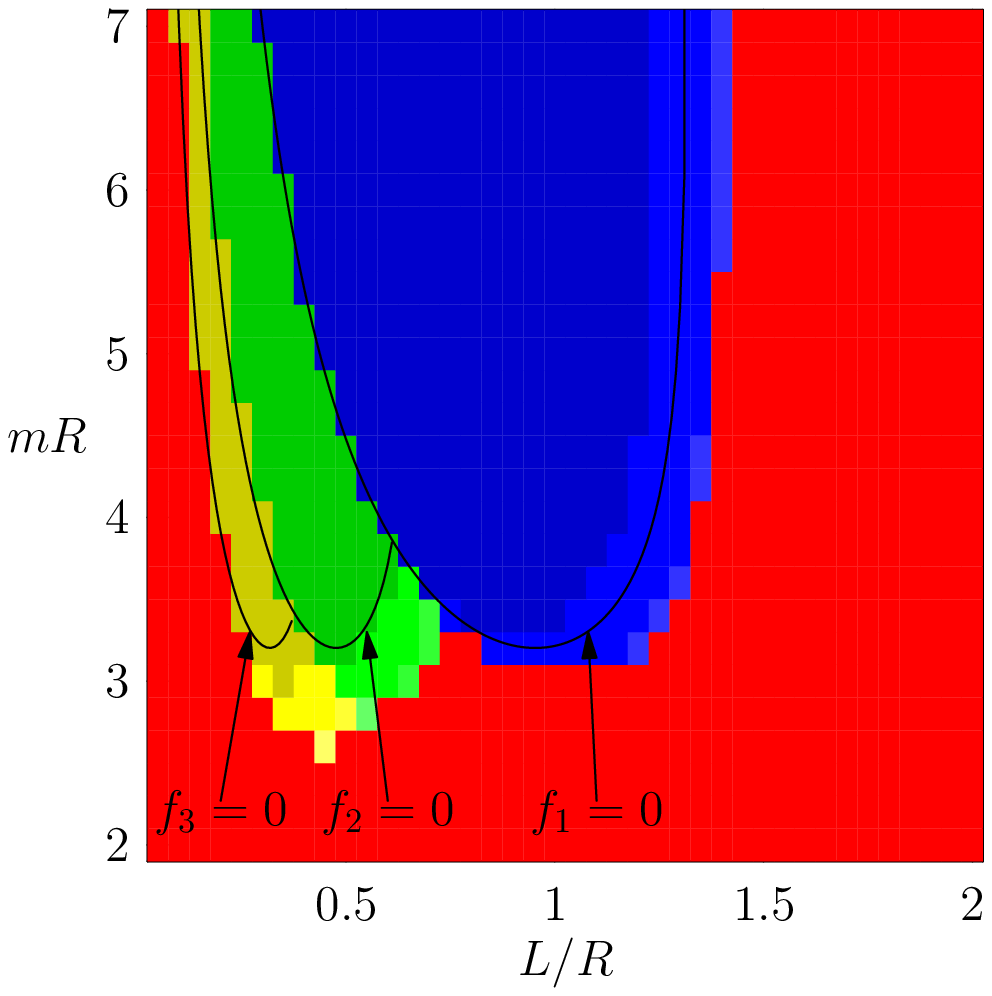}
    \end{center}
  \end{minipage}
  \hfill
  \caption{\small Phase diagram of QCD(Adj) for $N_f  = 2$ in the ($L / R$,
    $m R$) plane: (Top Left) $N = 3$: $2.6 < (m R)_{c \rightarrow d} <
    2.8$; (Top Right) $N = 4$: $2.8 < (m R)_{c \rightarrow d} < 3.0$;
    (Bottom Left) $N = 5$: $2.8 < (m R)_{c \rightarrow d} < 3.0$;
    (Bottom Right) $N = 6$: $3.0 < (m R)_{c \rightarrow d} < 3.2$. The $f_n = 0$ curves indicate the lines of transition for the $N = \infty$ result.} 
  \label{adj_nf2_mR}
\end{figure}

Consider
$SU(3)$ QCD(Adj). The phase diagram for $N_f  = 4$ Majorana
flavours in the $(L / R, m R)$ plane is shown in Figure \ref{adjn3nf4}
(Left), and in the $(L / R, m L)$ plane in Figure \ref{adjn3nf4}
(Right). The phases are defined according to the value of $\Tr
P$. The confined (red) phase has $\Tr P^k = 0$ for all $k$. The
2-gap (green) phase can be distinguished from the de-confined or 1-gap
(blue) phase in either of two ways: 1) The 2-gap phase has $\left| \Tr
  P^2 \right| > \left| \Tr P \right|$, whereas the de-confined phase
has $\left| \Tr P \right| > \left| \Tr P^2 \right|$, 2) The 2-gap
phase has ${\rm Proj}_{\Z_3} \Tr P < 0$ whereas the de-confined phase
has ${\rm Proj}_{\Z_3} \Tr P > 0$, where ${\rm Proj}_{\Z_3}$ indicates
projection onto the nearest $\Z_3$ axis. Darker shading in a $k$-gap
phase indicates a greater magnitude of $\left| \frac{1}{N} \Tr P^k
\right|$. The pure Yang-Mills theory transition is visible at large $m
R$ for $1.55 < (L / R)_{c \rightarrow d} < 1.6$ in approximate agreement with the
$N = 3$ result in Figure \ref{yang-mills}. For small enough $m R$
phase transitions are avoidable for all $L / R$. The critical value
$(m R)_\text{crit}$ below which $\Z_N$ symmetry-breaking phase transitions
are absent is less for smaller $N_f $, but the overall shape of the
phase diagram is otherwise qualitatively similar. 


\begin{figure}[t]
  \hfill
  \begin{minipage}[t]{.45\textwidth}
    \begin{center}  
      \includegraphics[width=0.95\textwidth]{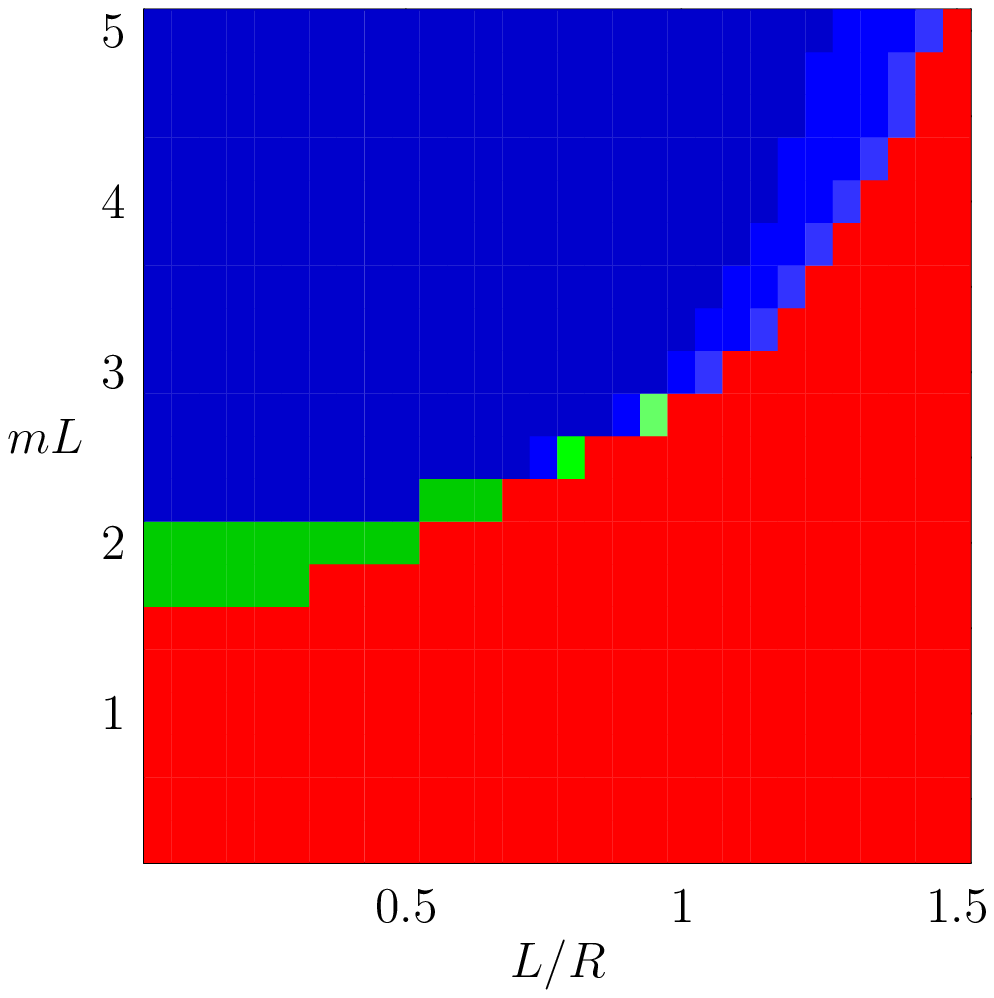}
    \end{center}
  \end{minipage}
  \hfill
  \begin{minipage}[t]{.45\textwidth}
    \begin{center}
\includegraphics[width=0.95\textwidth]{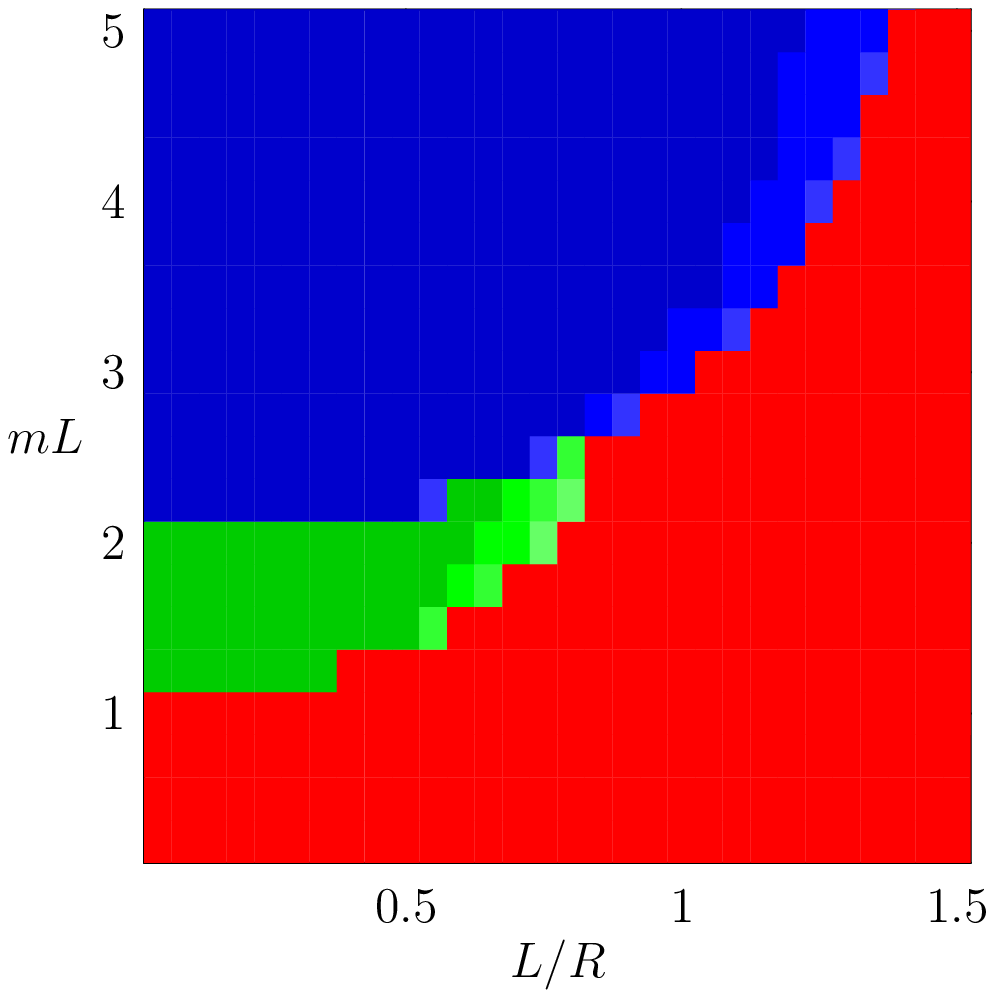}
    \end{center}
  \end{minipage}
  \hfill

  \hfill
  \begin{minipage}[t]{.45\textwidth}
    \begin{center}  
      \includegraphics[width=0.95\textwidth]{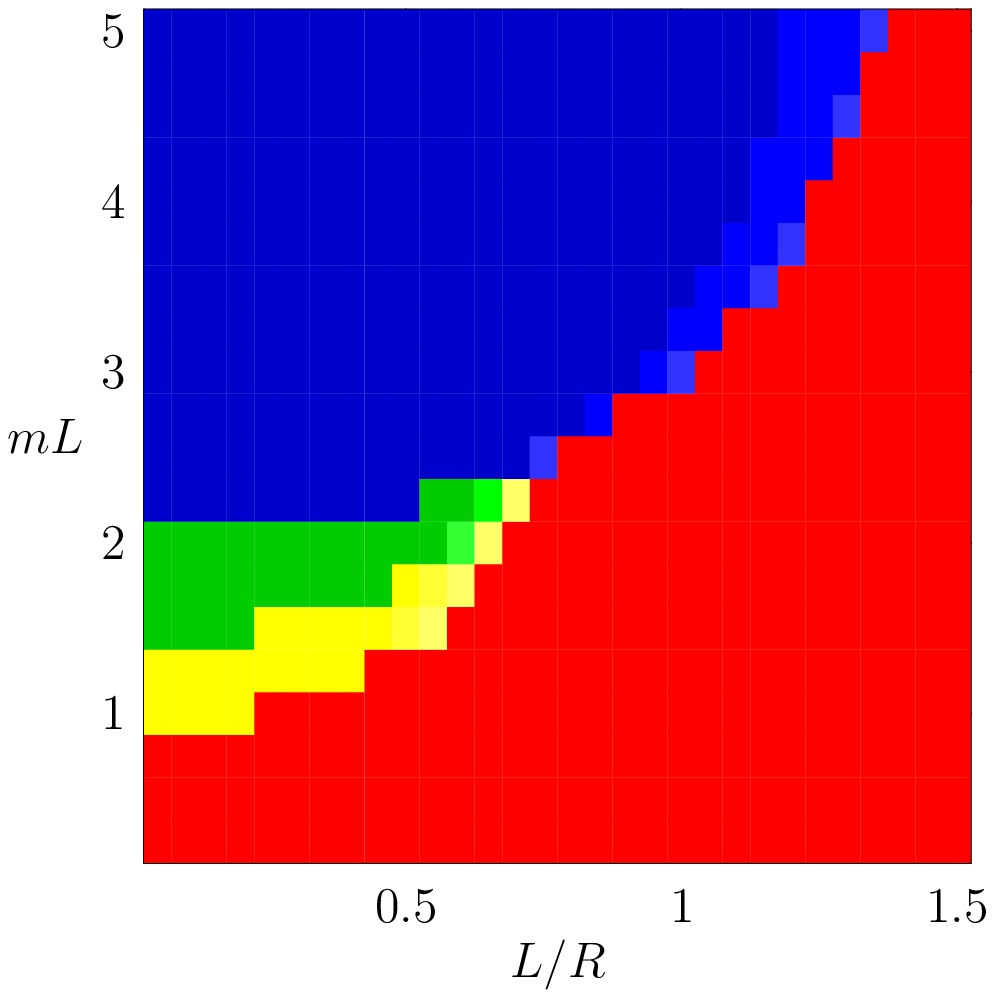}
    \end{center}
  \end{minipage}
  \hfill
  \begin{minipage}[t]{.45\textwidth}
    \begin{center}
\includegraphics[width=0.95\textwidth]{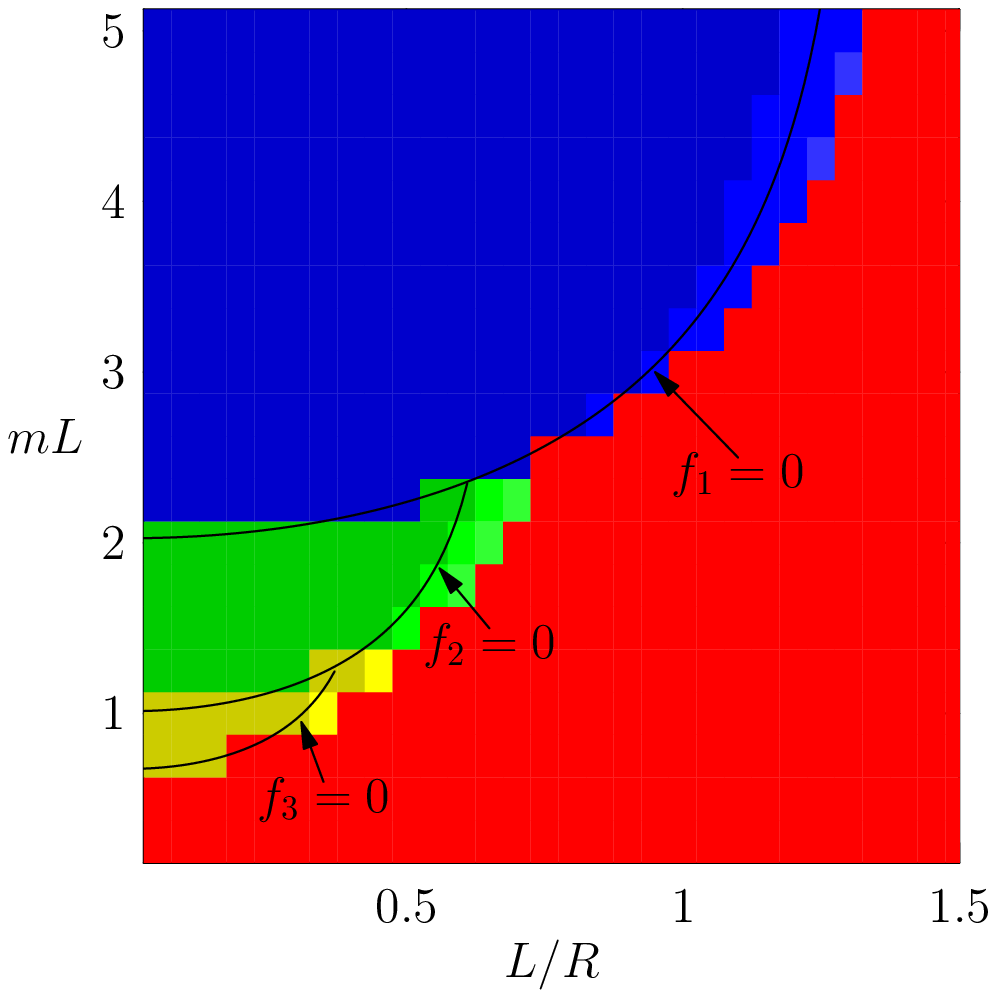}
    \end{center}
  \end{minipage}
  \hfill
  \caption{\small Phase diagram of QCD(Adj) for $N_f  = 2$ in the ($L / R$, $m
    L$) plane: (Top Left) $N = 3$: $1.5 < (m L)_c < 1.75$; (Top Right)
    $N = 4$: $1.0 < (m L)_c < 1.25$; (Bottom Left) $N = 5$: $0.75 < (m
    L)_c < 1.0$; (Bottom Right) $N = 6$: $0.5 < (m L)_c < 0.75$. The $f_n = 0$ curves indicate the lines of transition for the $N = \infty$ result.} 
  \label{adj_nf2_mb}
\end{figure}

To see how the large $N$ phase diagram unfolds it is useful to also
consider the phase diagrams for $N = 4, 5, 6$. To compare with the
${\field R}^3 \times S^1$ results in \cite{Myers:2009df} we consider
$N_f  = 2$. The phase diagrams of $N = 3, 4, 5, 6$ QCD(Adj) are shown
in the $(L / R, m R)$ plane in Figure \ref{adj_nf2_mR}. As $N$ is
increased a new  phase is formed for each odd N. For $N = 3$ this
phase is the 2-gap phase with $\left| \Tr P^2 \right| > \left| \Tr P
\right|$. For $N = 5$ the new phase is the 3-gap (yellow) phase with $\left|
  \Tr P^3 \right| > \left| \Tr P^2 \right|, \left| \Tr P \right|$. For
even $N$ the phases fan out into the region of small $L / R$. But, for
$N \ge 7$ the phase diagram gets even more complicated. For example,
from \cite{Myers:2009df} we know that for $N = 7$ it is possible to
have two different 3-gap phases, one which maximizes $\left| \Tr P^3
\right|$ and the other which maximizes $\left| \Tr P^4
\right|$. However, if $N \mod k = 0$ then there is always a k-gap
phase for which the eigenvalues are distributed in $k$ evenly spaced
clumps containing $N / k$ eigenvalues per clump such that $\Tr P^k \ne
0$ and $\Tr P^l = 0$ for $l \ne k$. If $N\mod k \ne 0$ then there is
still a k-gap phase with $k$ clumps of eigenvalues and for which
$\left| \Tr P^k \right| > \left| \Tr P^l \right|$ for $l$ not a
multiple of $k$. We suspect the other types of $k$-gap phases to
result as a consequence of the decreased symmetry of the finite $N$
theory, and conjecture that they should not be present in limit $N
\rightarrow \infty$.

As $N$ is increased the new phases extend down into lower values of
$(m R)$ than the de-confined phase, however when the next phase is
formed the previous phases are dragged up into regions of larger $(m
R)$. This is what we expect given that the critical mass in the $N
\rightarrow \infty$ limit occurs at $m R= 3.203$ for all the gapped
phases. This phenomenon is particularly clear by examination of $(m
R)_{c \rightarrow d}$ of the de-confined phase for $N = 3, 4, 5, 6$ in
Figure  \ref{adj_nf2_mR}, which shows a steady increase in $(m R)_{c
  \rightarrow d}$ from $2.6 < (m R)_{c \rightarrow d} < 2.8$ for $N =
3$ to $3.0 < (m R)_{c \rightarrow d} < 3.2$ for $N = 6$.

It serves as a useful check to compare quantitatively with the results on
${\field R}^3 \times S^1$ in \cite{Myers:2009df}. This can be done by
plotting the phase diagram in the $(L / R, m L)$ plane and considering
the limit $L / R \rightarrow 0$. The phase diagrams for $N = 3, 4, 5,
6$ in the $(L / R, m L)$ plane are shown in Figure
\ref{adj_nf2_mb}. Taking the limit $L / R \rightarrow 0$ in these
phase diagrams shows precise agreement with the ${\field R}^3 \times
S^1$ results in \cite{Myers:2009df} for the values of $m L$ at which
the transitions occur. Away from the limit of small $L / R$ the phases
do not take the precise Polyakov loop eigenvalues which are determined
for ${\field R}^3 \times S^1$ in \cite{Myers:2009df}, rather the
eigenvalues spread out slowly around the circle as $L / R$ is increased, causing the
magnitude $\left| \Tr P \right|$ to decrease. 

\subsection{Comparison with lattice results of Cossu and D'Elia}

\begin{figure}[t]
  \hfill
  \begin{minipage}[t]{.48\textwidth}
    \begin{center}  
      \includegraphics[width=0.95\textwidth]{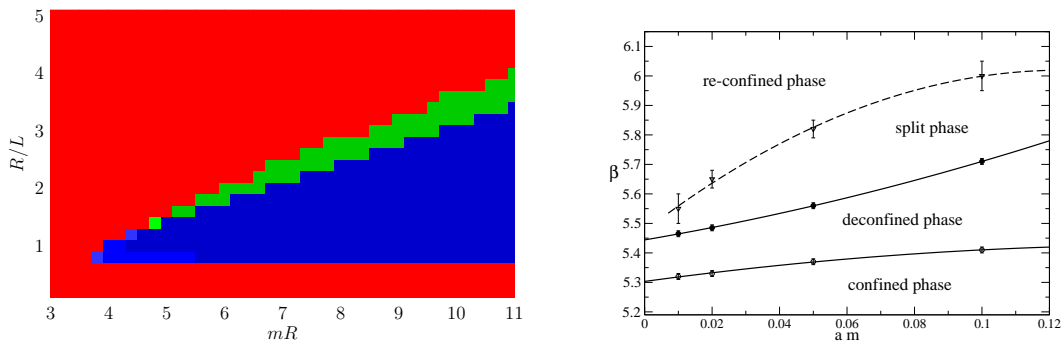}
    \end{center}
  \end{minipage}
  \hfill
  \begin{minipage}[t]{.42\textwidth}
    \begin{center}
\includegraphics[width=0.95\textwidth]{phdiagram.eps}
    \end{center}
  \end{minipage}
  \hfill
  \caption{\small QCD(Adj) for $N = 3$, $N_f  = 4$: (Left) ($m R$, $L /
    R$). $L = 2 \pi R_{S^1}$. Only the confined phase persists for $m
    R \lesssim 3.6$; (Right) Results from lattice simulations of Cossu
    and D'Elia \cite{Cossu:2009sq} on a $16^3 \times L_c$
    lattice. Here $\beta$ is related to the inverse coupling $\beta =
    2 N / g^2$.} 
  \label{latt_comp}
\end{figure}

To connect with strong coupling results it is useful to qualitatively
compare the phase diagram on $S^3 \times S^1$ to that from the recent lattice
simulations in \cite{Cossu:2009sq}. In \cite{Cossu:2009sq} the authors
also consider QCD(Adj) with periodic boundary conditions on
fermions for the purpose of checking the volume dependence of the
phase diagram. They obtain results for $N = 3$ and $N_f  = 4$ (or
$N_f ^D = 2$ Dirac flavours). To compare with their results we
calculated the phase diagram on $S^3 \times S^1$ in the $(m R, R / L)$
plane as shown in Figure \ref{latt_comp} (Left). The boundaries of the
2-gap phase correspond to the expected lines of transition in the
limit of large $R$, {\it i.e.\/}~the ${\field R}^3 \times S^1$ result. With
permission of the kind authors of \cite{Cossu:2009sq} we show their
phase diagram in the $(m a, \beta)$ plane in Figure \ref{latt_comp}
(Right), where $m$ is the fermion mass, $a$ is the lattice
spacing, and $\beta$ is the lattice parameter which goes like the
inverse coupling $\beta = 2 N / g^2$. It is important to note that
this quantity is frequently confused with $L$ (which is often also
called $\beta$), however the lattice $\beta = 2 N / g^2$ goes more
like $1/L$. Obtaining a more quantitative relationship between $\beta$
and $L$ would allow for an even better comparison than what we have
shown. 

In the lattice phase diagram of Figure \ref{latt_comp} (Right) it is
unclear whether or not the de-confined phase persists into the chiral
limit. In the perturbative phase diagram of Figure \ref{latt_comp}
(Left) it does not, however, it is possible that in the case of more
strongly interacting adjoint fermions the de-confined phase drops down
to lower values of $m R$. One thing which may help answer this
question is to plot the lattice phase diagram in the $(m L_s, L_s /
L_t)$ plane, as it may show a clearer trend. 

It is interesting to compare as well with the lattice phase diagram
which results from adding a single double trace term to the pure gauge
theory action \cite{Myers:2007vc,Wozar:2008nv}. The phase diagrams for
these theories seem to suggest that the confined phase passes through (i.e., the strong and weak-coupling confined phases are connected)
when the double trace term is strong enough. The double trace term,
which is approximately the contribution of static massive adjoint
fermions, is given by $h \Tr_\text{adj} P$ for $N = 3$, where larger magnitudes
of $h$ correspond to smaller fermion mass.\footnote{More information
  on double trace deformations can be found in
  \cite{Schaden:2005fs,Ogilvie:2007tj,Unsal:2008ch}.} 

\begin{figure}[t]
  \hfill
  \begin{minipage}[t]{.48\textwidth}
    \begin{center}  
      \includegraphics[width=0.95\textwidth]{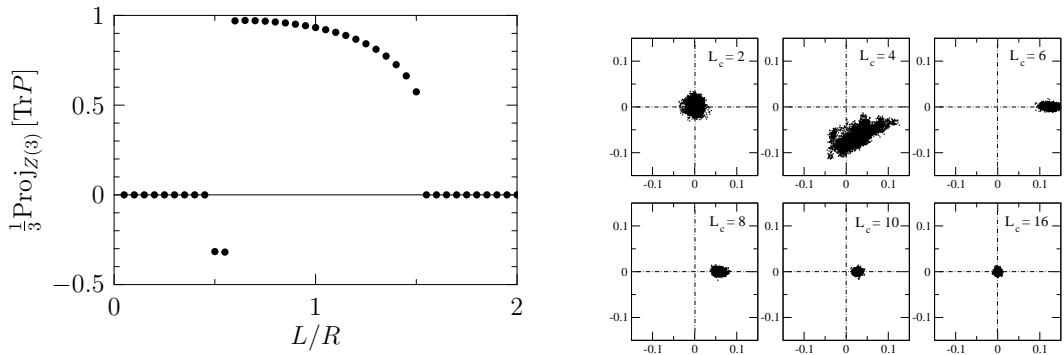}
    \end{center}
  \end{minipage}
  \hfill
  \begin{minipage}[t]{.42\textwidth}
    \begin{center}
\includegraphics[width=0.95\textwidth]{UVfixed.eps}
    \end{center}
  \end{minipage}
  \hfill
  \caption{\small QCD(Adj) for $N = 3$, $N_f  = 4$: (Left) Results from
    perturbation theory in ($L / R$, $\frac{1}{3} {\rm Proj}_{\Z_3}
    \Tr P$) plane. $(m R) = 6$; (Right) Results from lattice
    simulations of Cossu and D'Elia \cite{Cossu:2009sq} in the $({\rm
      Re} \left[ \Tr P \right], {\rm Im} \left[ \Tr P \right] )$ plane
    on $16^3 \times L_c$ lattices for $\beta = 5.75$ and $a m =
    0.10$.} 
  \label{latt_trp}
\end{figure}

Another remarkable similarity between the $S^3 \times S^1$ result and
the lattice results \cite{Cossu:2009sq} of Cossu and D'Elia is that
there is good qualitative agreement for $\Tr P$ as a function of $L$
for a fixed value of the fermion mass. In Figure \ref{latt_trp} we
show $\Tr P$ as a function of the length of the temporal
direction. The $S^3 \times S^1$ result in the $(L / R, \frac{1}{3}
{\rm Proj}_{\Z_3} \Tr P)$ plane is in Figure \ref{latt_trp}
(Left). The lattice result of \cite{Cossu:2009sq} in the $({\rm Re}
\left[ \Tr P \right], {\rm Im} \left[ \Tr P \right])$ plane for $16^3
\times L_c$ lattices with increasing length $L_c$ of the temporal dimension is given in Figure \ref{latt_trp} (Right). What is remarkable
about this comparison is that the perturbative result appears to
capture even some fine details of the phase diagram obtained on the
lattice including microscopic changes in the magnitude of $\Tr P$. The
sharp discontinuity in $\Tr P$ at the transition to and from the 2-gap
phase, and the slow drop in $\Tr P$ in the de-confined phase as the
confined phase is approached, seem to agree rather well. 

It might, at first, seem surprising that agreement between the lattice
calculation on $(S^1)^4$ and our calculation on $S^3 \times S^1$ is so good since
these spaces have different first homotopy groups. However, 3 of the circles
of the torus in the lattice calculation are large and so one might expect
that the one remaining small circle plays the dominant role in determining
the phase structure \footnote{We wish to thank Mithat Unsal for useful discussions on this topic.}. In addition, it seems that the phase diagram doesn't change much as the
coupling strength is increased from the weakly interacting limit to
the strongly interacting limit when considering the phase diagram in
terms of patterns of $\Z_N$ breaking. This is not necessarily the case for all observables.

\section{Discussion}

It is important to clarify the implications of our results for the
issue of volume independence. The latter relies on the fact that the
theory is in the confining phase with unbroken centre symmetry. Our
result show that at weak coupling on $S^3\times S^1$ the confining
phase persists in the limit $L/R\to0$ as long as the fermion mass is below a
critical value in units of $1/R$, {\it i.e.\/}~$1.225/R$ for
$N_f=1$, ~$3.203/R$ for
$N_f=2$. The critical mass increases with increasing $N_f$.
Of course, our result is valid on a small $S^3$ and the
interesting question is what happens to the critical mass as one moves
to strong coupling by taking $R>1/\Lambda_\text{QCD}$. 

The general analysis presented in this paper could just as well be
applied to 
many other theories. For example, it is straightforward to consider
QCD(Adj) with anti-periodic boundary conditions which is of interest
since lattice simulations have been done
\cite{Hietanen:2008mr,DelDebbio:2008zf}, as well as analytical
calculations \cite{Poppitz:2009uq}, which suggest the presence of a
conformal window. It would also be interesting to consider symmetric
and antisymmetric representation fermions and compare with related
lattice results \cite{DeGrand:2008kx}. Interesting also, would be
applications to softly broken ${\cal N}=4$ theory, since there
are several fermions and scalars and more complicated mass hierarchies are
possible leading to more complicated phase diagrams. Whether a
connection could be made with the string theory dual remains to be
seen. In particular, it would be interesting to understand how the
confinement/de-confinement transitions can occur with periodic boundary
conditions for the fermions suggesting that it is not a Hawking-Page
transition in AdS. The other related issue is how the nature of the
transition changes as interactions are turned on and one moves to
strong coupling. As we mentioned in the introduction, for pure gauge
theory the 3-loop calculation in \cite{Aharony:2005bq} shows that the
transition is first order and occurs at a lower termperature, {\it
  i.e.\/}~larger $L/R$, than the
non-interacting Hagedorn transition. Unfortunately it will be very
difficult to generalize the calculation of \cite{Aharony:2005bq} to
include massive fermions.

Calculations on $S^3 \times
S^1$ can be used to define better the extent of perturbative validity
of orientifold planar equivalence \cite{Armoni:2003gp,Armoni:2004ub},
which is a large $N$ equivalence of QCD(Adj) and QCD with
symmetric/antisymmetric representation fermions. The same can be done
for orbifold planar equivalence \cite{Kovtun:2003hr,Kovtun:2004bz},
the equivalence of QCD(Adj) and QCD with bi-fundamental
representation fermions. In both cases a comparison of large $N$ phase
diagrams is possible, and perhaps some observables could be
compared. The one-loop effective potential and phase diagrams for
QCD(Adj/AS/S) on $S^3 \times S^1$ with massless fermions were computed
in the very clearly written papers
\cite{Hollowood:2006cq,Unsal:2007fb}. One could also study QCD
(fundamental representation fermions) and incorporate a finite
chemical potential. However, in this case one has to confront
the sign problem. It would be interesting to 
compare a phase-quenched result (demanding a real fermion determinant) on $S^3 \times S^1$ with results from the several different
techniques for dealing with the sign problem in QCD at various coupling strengths,
a diverse sampling of which can be
found in
\cite{Aarts:2008rr,Hands:2006ve,D'Elia:2007ke,deForcrand:2006pv,Akemann:2004dr,Alford:2007xm,Ogilvie:2008zt}. 

There are several things that might be done to allow for better
comparison of weak-coupling results on $S^3 \times S^1$ to lattice
results. Using the two-loop renormalization group equation to give a
fitting function for the relationship between the lattice parameter
$\beta = 2 N / g^2$ and the length $L$ of $S^1$ would allow for more
quantitative comparisons. More lattice results on different volumes
and for various $N$ would show if phase diagrams are consistently
similar. In particular, we see a pattern emerging that suggests that
for QCD(Adj) with finite even $N$ there are $N / 2$ gapped phases
(including the de-confined phase) with the property that $\left| \Tr
  P^k \right| > \left| \Tr P^l \right|$ in a k-gap phase for $l$ not a
multiple of $k$, and for $N$ odd there are $(N+1)/2$ gapped phases
with this property (narrow regions of additional phases are also
possible). On the analytical side interactions might be included in
the weak-coupling effective potential by working out higher loop
orders. In addition, to compare even better with lattice results the
theory can be put on the torus. The one-loop effective potential for
QCD(Adj) with massive fermions was computed on ${\field R}^d \times
T^n$ in \cite{Barbon:2006us} (see also \cite{vanBaal:2000zc}). 

\acknowledgments

\noindent Our understanding of the presented results relies heavily on
relevant discussions with several colleagues. We would like to thank
Barak Bringoltz, Massimo D'Elia, Ari Hietanen,  Biagio Lucini,
Agostino Patella, and Antonio Rago for helpful discussions on
comparing with lattice results. In particular we would like to thank
Guido Cossu and Massimo D'Elia for allowing us to use their figures in
this paper, and to thank Massimo D'Elia for discussions of their
lattice results. We would also like to thank Barak Bringoltz and Ari
Hietanen for discussing their related lattice results. We are grateful
to Biagio Lucini and Antonio Rago for discussions on the analysis of
phase transitions. We are also grateful to Adi Armoni, Michael
Ogilvie, and Mithat {\" U}nsal for discussions on the implications to
volume independence, and to Carlos Hoyos, for discussing his related
calculation on the torus. Many of these discussions took place during
the fruitful Large $N$ conference in Swansea, during which progress
was made towards the completion of this work. 
TJH would like to acknowledge the support of STFC grant.
ST/G000506/1.

\startappendix

\Appendix{Spherical Harmonics}

In this appendix we collect together results for the spectra of
various Laplace operators on a sphere $S^d$. 

First of all the scalar Laplace equation is solved by generalized
spherical harmonics,
\begin{equation}
\Delta^{(s)} Y_{\ell,\vec m}(\hat\Omega) = - 
\varepsilon_l^{(s)2} Y_{\ell,\vec m}(\hat\Omega)\ .
\end{equation}
The eigenvalues are
\begin{equation}
\varepsilon_l^{(s)2} = \ell (\ell+d-1) R^{-2}\  ,
\end{equation}
and the degeneracy is
\begin{equation}
d_\ell^{(s)} = \frac{(2 \ell + d - 1) (\ell + d - 2) !}{\ell ! (d - 1) !} ,
\end{equation}
where the angular momentum $\ell=0,1,\ldots$.

Following \cite{DeNardo:1996kp} the eigenvalues of the vector
Laplacian $\Delta^{(v)}$ on vector fields are obtained by separation
of the longitudinal (L) and transverse (T) components. The spatial
gauge field is thus decomposed $A_i = B_i + C_i$, where $B_i$ is the
transverse component with $\nabla_i B_i = 0$, and $C_i$ is the
longitudinal component with $C^i = \nabla^i\chi$. 

The vector Laplacian acting on the longitudinal component $C_i$ has
the same spectrum as for the scalar Laplacian because 
\begin{equation}
\left( \nabla^i \nabla_i \delta^j_k - R^j{}_k \right) \nabla_j \chi =
\nabla_k \left( \nabla^i \nabla_i \chi\right)\ ,
\end{equation}
except that the $\ell=0$ mode is missing:
\begin{equation}
\varepsilon_\ell^{(v,L) 2} = \ell (\ell+d-1) R^{-2} ,
\end{equation}
for $\ell=1,2,\ldots$, and 
\begin{equation}
d_\ell^{(v,L)} = \frac{(2 \ell + d - 1) (\ell + d - 2) !}{\ell ! (d - 1) !} .
\end{equation}

The eigenvalues and degeneracy of the vector 
Laplacian on the transverse components $B_i$ are 
\begin{equation}
\varepsilon_\ell^{(v,T) 2} = \big( \ell (\ell + d - 1) + d - 2 \big)R^{-2}\ ,
\end{equation}
and
\begin{equation}
d_\ell^{(v,T)} = \frac{\ell (\ell + d - 1) (2\ell + d - 1) 
(\ell + d - 3) !}{(d - 2) ! (\ell + 1)!}\ .
\end{equation}
for $\ell=1,2,\ldots$.

The Laplacian on spinors is given by
\begin{equation}
\Delta^{(f)}=
\gamma^i \gamma^j \nabla_i \nabla_j = 
g^{i j} \nabla_i \nabla_j - \tfrac{1}{4}R\ ,
\end{equation}
where ${\cal R}$ is the scalar curvature of the sphere and 
\begin{equation}
\nabla_i = \partial_i + \Gamma_i
\end{equation}
is the covariant derivative on spinors with spin connection $\Gamma_i$. The eigenvalues and degeneracy
\cite{Candelas:1983ae,Camporesi:1992tm,Camporesi:1995fb} are
\EQ{
\varepsilon_l^{(f) 2} = \big( \ell + \tfrac{d}{2} \big)^2 R^{-2}\ ,
}
and
\EQ{
d_\ell^{(f)} = \frac{2 \big( d + \ell - 2 \big) !}{(\ell-1) ! (d - 1)
!}\ ,
}
where $\ell=1,2,\ldots$. 

\clearpage
\Appendix{The Abel-Plana Formula}

\begin{figure}[ht] 
\centerline{\includegraphics[width=2.5in]{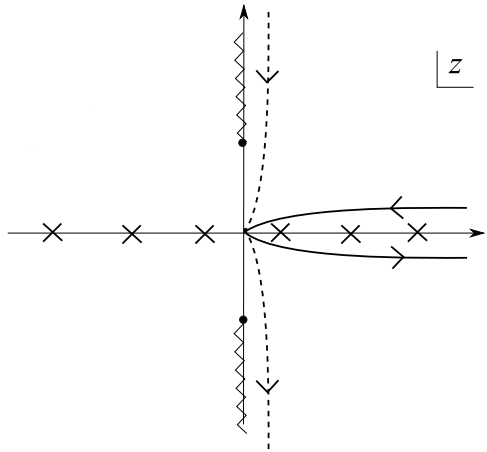}}
\caption{\small The contour used in the derivation of the
  Abel-Plana formula.}\label{fig2}
\end{figure}

In this appendix, we prove the version of the Abel-Plana formula
quoted in the main text \eqref{abelp}. The idea is to represent the
sum on left-hand side as a contour integral:
\EQ{
\sum_{\ell=0}^{\infty} f(\ell+\tfrac12)
=\frac i2\int_{{\cal C}}dz\,f(z)\tan(\pi z)\ ,
}
where ${\cal C}$ is the contour illustrated in Fig.~\ref{fig2}. In our
case, the function $f(z)$ has square root branch points at $z=\pm
imR$. Using the above we have
\EQ{
\sum_{\ell=0}^{\infty} f(\ell+\tfrac12)-\int_0^\infty dx\,f(x)=
\frac 1{2i}\int_0^{\infty+i\epsilon}dz\,f(z)\big(\tan(\pi z)-i\big)
-\text{c.c.}
}
Now we can rotate the contour here that runs from the origin out to
infinity over the poles so that it runs from the origin to $i\infty$
to the right of the branch point of $f(z)$ at $imR$. This gives the
right-hand side as
\EQ{
-i\int_0^{\infty}dx\,\frac{f(ix+\epsilon)}{e^{2\pi x}+1}-\text{c.c.}
}
Hence, we have proved that
\EQ{
\sum_{\ell=0}^{\infty} f(\ell+\tfrac12) = \int_0^{\infty} dx\,
f(x) - i \int_0^{\infty}dx \, \frac{f(i x +
  \epsilon) - f(-i x - \epsilon)}{e^{2 \pi x} + 1}\ . 
\label{abelpa}
}
which is the formula \eqref{abelp} in the text.

\clearpage
\Appendix{Numerical minimization}
\label{nmin}

In the finite $N$ calculations of the phase diagram it was necessary
to numerically minimize the effective potential. This was not trivial
as it is difficult to find an algorithm that will always find the
global minimum of an arbitrary function. Two techniques which proved
most useful were the Random Search and Differential Evolution
numerical minimization routines implemented in Mathematica, which are
reviewed in \cite{Champion:2002wr}. Increasing the number of search
points improves the chances of obtaining the global minimum. Somewhat
surprisingly in most cases just 5 search points were enough to obtain
the minimum accurate to around $13$ digits for $N = 3$. It is only
slightly less accurate when considering $N = 4, 5, 6$. The addition of
the mass term to the effective potential actually makes the
minimization easier. The pure gauge theory plots required many more
search points, between 500 and 1500. 

It was not possible to obtain reliable minimization of the effective
potential when there were terms represented as an infinite
series. Therefore it was necessary to put all infinite series in a
non-series expression, either by solving them, or converting them into
integral forms using the Abel-Plana formula.

As this type of calculation doesn't have much history (however, the
same procedure was used in \cite{Myers:2009df}) there is a lot of room
for improvement in technique.

It is additionally important to perform checks of the saddle point approximation by plotting the relevant observables as a function of the configuration space of the $\theta_i$ as discussed in the finite $N$ section.

\thebibliography{99}

{\small 

\bibitem{Aharony:2003sx}
  O.~Aharony, J.~Marsano, S.~Minwalla, K.~Papadodimas and M.~Van Raamsdonk,
  Adv.\ Theor.\ Math.\ Phys.\  {\bf 8}, 603 (2004)
  [arXiv:hep-th/0310285].
  
\bibitem{Aharony:2005bq}
  O.~Aharony, J.~Marsano, S.~Minwalla, K.~Papadodimas and M.~Van Raamsdonk,
  Phys.\ Rev.\  D {\bf 71} (2005) 125018
  [arXiv:hep-th/0502149].
  
\bibitem{Hallin:1998km}
  J.~Hallin and D.~Persson,
  Phys.\ Lett.\  B {\bf 429} (1998) 232
  [arXiv:hep-ph/9803234].
  
\bibitem{Sundborg:1999ue}
  B.~Sundborg,
  Nucl.\ Phys.\  B {\bf 573} (2000) 349
  [arXiv:hep-th/9908001].

\bibitem{Witten:1998zw}
  E.~Witten,
  Adv.\ Theor.\ Math.\ Phys.\  {\bf 2} (1998) 505
  [arXiv:hep-th/9803131].

\bibitem{Myers:2007vc}
  J.~C.~Myers and M.~C.~Ogilvie,
  Phys.\ Rev.\  D {\bf 77}, 125030 (2008)
  [arXiv:0707.1869 [hep-lat]].
  
\bibitem{Wozar:2008nv}
  C.~Wozar, T.~Kastner, B.~H.~Wellegehausen, A.~Wipf and T.~Heinzl,
  arXiv:0808.4046 [hep-lat].

\bibitem{Myers:2009df}
  J.~C.~Myers and M.~C.~Ogilvie,
  arXiv:0903.4638 [hep-th].

\bibitem{Bedaque:2009md}
  P.~F.~Bedaque, M.~I.~Buchoff, A.~Cherman and R.~P.~Springer,
  arXiv:0904.0277 [hep-th].

\bibitem{Cossu:2009sq}
  G.~Cossu and M.~D'Elia,
  arXiv:0904.1353 [hep-lat].

\bibitem{Bringoltz:2009mi}
  B.~Bringoltz,
  arXiv:0905.2406 [hep-lat].

\bibitem{Bringoltz:2009kb}
  B.~Bringoltz and S.~R.~Sharpe,
  arXiv:0906.3538 [hep-lat].

\bibitem{Kovtun:2007py}
  P.~Kovtun, M.~Unsal and L.~G.~Yaffe,
  JHEP {\bf 0706} (2007) 019
  [arXiv:hep-th/0702021].
  
\bibitem{Boyd:1996bx}
  G.~Boyd, J.~Engels, F.~Karsch, E.~Laermann, C.~Legeland, M.~Lutgemeier and B.~Petersson,
  Nucl.\ Phys.\  B {\bf 469} (1996) 419
  [arXiv:hep-lat/9602007].
  
\bibitem{Eguchi:1982nm}
  T.~Eguchi and H.~Kawai,
  Phys.\ Rev.\ Lett.\  {\bf 48}, 1063 (1982).
  
\bibitem{Yaffe:1981vf}
  L.~G.~Yaffe,
  Rev.\ Mod.\ Phys.\  {\bf 54} (1982) 407.

\bibitem{Bhanot:1982sh}
  G.~Bhanot, U.~M.~Heller and H.~Neuberger,
  Phys.\ Lett.\  B {\bf 113} (1982) 47.
  
\bibitem{Kazakov:1982gh}
  V.~A.~Kazakov and A.~A.~Migdal,
  Phys.\ Lett.\  B {\bf 116} (1982) 423.
  
\bibitem{Okawa:1982ic}
  M.~Okawa,
  Phys.\ Rev.\ Lett.\  {\bf 49} (1982) 353.
  
\bibitem{Bhanot:1982sh}
  G.~Bhanot, U.~M.~Heller and H.~Neuberger,
  Phys.\ Lett.\  B {\bf 113} (1982) 47.
  
\bibitem{Bringoltz:2008av}
  B.~Bringoltz and S.~R.~Sharpe,
  Phys.\ Rev.\  D {\bf 78} (2008) 034507
  [arXiv:0805.2146 [hep-lat]].
  
\bibitem{GonzalezArroyo:1982ub}
  A.~Gonzalez-Arroyo and M.~Okawa,
  Phys.\ Lett.\  B {\bf 120} (1983) 174.
  
\bibitem{Teper:2006sp}
  M.~Teper and H.~Vairinhos,
  Phys.\ Lett.\  B {\bf 652} (2007) 359
  [arXiv:hep-th/0612097].
  
\bibitem{Azeyanagi:2007su}
  T.~Azeyanagi, M.~Hanada, T.~Hirata and T.~Ishikawa,
  JHEP {\bf 0801} (2008) 025
  [arXiv:0711.1925 [hep-lat]].
  
\bibitem{Hanada:2009kz}
  M.~Hanada, L.~Mannelli and Y.~Matsuo,
  arXiv:0905.2995 [hep-th].
  
\bibitem{Ishiki:2009sg}
  G.~Ishiki, S.~W.~Kim, J.~Nishimura and A.~Tsuchiya,
  arXiv:0907.1488 [hep-th].
  
\bibitem{Hollowood:2006cq}
  T.~J.~Hollowood and A.~Naqvi,
  JHEP {\bf 0704}, 087 (2007)
  [arXiv:hep-th/0609203].
  
\bibitem{Hollowood:2006xb}
  T.~Hollowood, S.~P.~Kumar and A.~Naqvi,
  JHEP {\bf 0701}, 001 (2007)
  [arXiv:hep-th/0607111].
  
\bibitem{Hollowood:2008gp}
  T.~J.~Hollowood, S.~P.~Kumar, A.~Naqvi and P.~Wild,
  JHEP {\bf 0808} (2008) 046
  [arXiv:0803.2822 [hep-th]].

\bibitem{Lucini:2005vg}
  B.~Lucini, M.~Teper and U.~Wenger,
  JHEP {\bf 0502} (2005) 033
  [arXiv:hep-lat/0502003].
  
\bibitem{Schaden:2005fs}
  M.~Schaden,
  Nucl.\ Phys.\ Proc.\ Suppl.\  {\bf 161} (2006) 210
  [arXiv:hep-th/0511046].
\bibitem{Ogilvie:2007tj}
  M.~C.~Ogilvie, P.~N.~Meisinger and J.~C.~Myers,
  PoS {\bf LAT2007} (2007) 213
  [arXiv:0710.0649 [hep-lat]].
  
\bibitem{Unsal:2008ch}
  M.~Unsal and L.~G.~Yaffe,
  Phys.\ Rev.\  D {\bf 78} (2008) 065035
  [arXiv:0803.0344 [hep-th]].
  
\bibitem{Hietanen:2008mr}
  A.~J.~Hietanen, J.~Rantaharju, K.~Rummukainen and K.~Tuominen,
  JHEP {\bf 0905} (2009) 025
  [arXiv:0812.1467 [hep-lat]].
  
\bibitem{DelDebbio:2008zf}
  L.~Del Debbio, A.~Patella and C.~Pica,
  arXiv:0805.2058 [hep-lat].

\bibitem{Poppitz:2009uq}
  E.~Poppitz and M.~Unsal,
  arXiv:0906.5156 [hep-th].

\bibitem{DeGrand:2008kx}
  T.~DeGrand, Y.~Shamir and B.~Svetitsky,
  Phys.\ Rev.\  D {\bf 79} (2009) 034501
  [arXiv:0812.1427 [hep-lat]].
\bibitem{Armoni:2003gp}
  A.~Armoni, M.~Shifman and G.~Veneziano,
  Nucl.\ Phys.\  B {\bf 667} (2003) 170
  [arXiv:hep-th/0302163].
  
\bibitem{Armoni:2004ub}
  A.~Armoni, M.~Shifman and G.~Veneziano,
  Phys.\ Rev.\  D {\bf 71} (2005) 045015
  [arXiv:hep-th/0412203].
  
\bibitem{Kovtun:2003hr}
  P.~Kovtun, M.~Unsal and L.~G.~Yaffe,
  JHEP {\bf 0312} (2003) 034
  [arXiv:hep-th/0311098].
  
\bibitem{Kovtun:2004bz}
  P.~Kovtun, M.~Unsal and L.~G.~Yaffe,
  JHEP {\bf 0507} (2005) 008
  [arXiv:hep-th/0411177].
 
\bibitem{Unsal:2007fb}
  M.~Unsal,
  Phys.\ Rev.\  D {\bf 76}, 025015 (2007)
  [arXiv:hep-th/0703025].

\bibitem{Aarts:2008rr}
  G.~Aarts and I.~O.~Stamatescu,
  JHEP {\bf 0809} (2008) 018
  [arXiv:0807.1597 [hep-lat]].
  
\bibitem{Hands:2006ve}
  S.~Hands, S.~Kim and J.~I.~Skullerud,
  Eur.\ Phys.\ J.\  C {\bf 48} (2006) 193
  [arXiv:hep-lat/0604004].
  
\bibitem{D'Elia:2007ke}
  M.~D'Elia, F.~Di Renzo and M.~P.~Lombardo,
  Phys.\ Rev.\  D {\bf 76} (2007) 114509
  [arXiv:0705.3814 [hep-lat]].
  
\bibitem{deForcrand:2006pv}
  P.~de Forcrand and O.~Philipsen,
  JHEP {\bf 0701} (2007) 077
  [arXiv:hep-lat/0607017].
  
\bibitem{Akemann:2004dr}
  G.~Akemann, J.~C.~Osborn, K.~Splittorff and J.~J.~M.~Verbaarschot,
  Nucl.\ Phys.\  B {\bf 712} (2005) 287
  [arXiv:hep-th/0411030].
  
\bibitem{Alford:2007xm}
  M.~G.~Alford, A.~Schmitt, K.~Rajagopal and T.~Schafer,
  Rev.\ Mod.\ Phys.\  {\bf 80} (2008) 1455
  [arXiv:0709.4635 [hep-ph]].
  
\bibitem{Ogilvie:2008zt}
  M.~C.~Ogilvie and P.~N.~Meisinger,
  arXiv:0812.0176 [hep-th].

\bibitem{Barbon:2006us}
  J.~L.~F.~Barbon and C.~Hoyos-Badajoz,
  Phys.\ Rev.\  D {\bf 73}, 126002 (2006)
  [arXiv:hep-th/0602285].

\bibitem{vanBaal:2000zc}
  P.~van Baal,
  arXiv:hep-ph/0008206.

\bibitem{DeNardo:1996kp}
  L.~De Nardo, D.~V.~Fursaev and G.~Miele,
  Class.\ Quant.\ Grav.\  {\bf 14}, 1059 (1997)
  [arXiv:hep-th/9610011].

\bibitem{Candelas:1983ae}
  P.~Candelas and S.~Weinberg,
  Nucl.\ Phys.\  B {\bf 237}, 397 (1984).
  
\bibitem{Camporesi:1992tm}
  R.~Camporesi,
  Commun.\ Math.\ Phys.\  {\bf 148} (1992) 283.
  
\bibitem{Camporesi:1995fb}
  R.~Camporesi and A.~Higuchi,
  J.\ Geom.\ Phys.\  {\bf 20}, 1 (1996)
  [arXiv:gr-qc/9505009].
  
\bibitem{Champion:2002wr}
  B.~Champion,
  ``Numerical Optimization in Mathematica: An Insider's View of NMinimize,''
  2002 World Multiconference on Systemics, Cybernetics, and Informatics (SCI 2002) Proceedings,
  http://library.wolfram.com/infocenter/Conferences/4311/
}
\end{document}